\documentclass[superscriptaddress, twocolumn, prx, aps, longbibliography, nofootinbib]{revtex4-2}
\usepackage{tikz}
\usepackage{amsmath, amssymb, color, wasysym, esint, bm}
\usepackage{graphicx}
\usepackage{booktabs, multirow}
\usepackage{float}
\usepackage{lineno}
\definecolor{linkcolor}{rgb}{0,0,0.6} 
\usepackage[pdftex,colorlinks=true,
	pdfstartview = FitV,
	linkcolor    = linkcolor,
	citecolor    = linkcolor,
	urlcolor     = linkcolor,
	hyperindex   = true,
	hyperfigures = false]{hyperref}

\begin{document}

\title{Active pistons extract work by periodic compression alone}

\author{Paul Bernard}
\affiliation{Gulliver UMR CNRS 7083, ESPCI Paris, PSL Research University, 10 rue Vauquelin, 75005 Paris, France}

\author{Euijoon Kwon}
\affiliation{Quantum Universe Center, Korea Institute for Advanced Study, Seoul 02455, Republic of Korea}

\author{Benjamin Buhl}
\affiliation{Gulliver UMR CNRS 7083, ESPCI Paris, PSL Research University, 10 rue Vauquelin, 75005 Paris, France}
\affiliation{Institut Lumière Matière UMR5306 - UCBL - CNRS,
10 rue Ada Byron, 69622 Villeurbanne, France}

\author{\'Etienne Fodor}
\affiliation{Department of Physics and Materials Science, University of Luxembourg, L-1511 Luxembourg City, Luxembourg}

\author{Olivier Dauchot}
\affiliation{Gulliver UMR CNRS 7083, ESPCI Paris, PSL Research University, 10 rue Vauquelin, 75005 Paris, France}

\begin{abstract}
Active matter is liable to invent protocols that evade the constraints of equilibrium thermodynamics. We put forward active pistons that extract work by periodic compression alone without changing any bulk property of the system. Such pistons necessarily couple the perturbation imposed by an external operator with some degrees of freedom internal to active components. We illustrate this design principle with elastic networks composed of self-aligning motile particles. For slow protocols, self-alignment always overwhelms mechanical friction when the internal activity exceeds a specific threshold controlled by fluctuations. We identify the key response coefficient that helps delineate regimes of work extraction, and reveal that the corresponding phase diagram follows a master curve with re-entrance in terms of noise amplitude. Overall, our active pistons embody a novel design principle with broad implications for building innovative engines far from equilibrium.
\end{abstract}

\maketitle


The study of cyclic engines has been pivotal in establishing the foundations of equilibrium thermodynamics~\cite{Carnot}. Carnot and Stirling, among others, have offered explicit protocols that extract work from cycles. Examining the performance of these minimal, yet non-trivial engines has led to the formulation of generic principles that guide the design of more complex engines. Chief among them is the necessity to vary temperature when operating cycles with thermal systems~\cite{Seifert2012, Bechinger2012, Parrondo2016}.

The study of engines now extends to active liquids~\cite{Sood2016, Cates2021, Pietzonka2022} where autonomous self-propelled agents sustain their dynamics far from equilibrium~\cite{Marchetti2013, Bechinger2016}. These liquids feature anomalous thermo-mechanical properties~\cite{Solon2015, Junot2017, Tociu2019, Bertin2019} that have been leveraged to design cycles at constant temperature~\cite{Ekeh2020, Holubec2020, Stark2021, Speck2022} and ratchet mechanisms~\cite{Prost1999, Leonardo2010, Cates2016, Pietzonka2019, Tailleur2026}. Since monothermal cycles cannot operate with thermal systems, these cycles specifically probe how active matter deviates from equilibrium~\cite{Nardini2016, Jack2022, Tailleur2022}. While any nonequilibrium system is prone to monothermal work extraction, the real challenge is to offer concrete protocols that demonstrate how active matter is particularly appropriate: it naturally stores an energy resource (e.g., autonomous self-propulsion) available for advantageous practical use (i.e., macroscopic extracted work).

Here, we put forward a novel class of active engines that operate by oscillating a single control parameter: we design {\em active pistons} that only require periodic compression, without changing any parameter other than volume, to extract work. To this end, we exploit self-alignment, an intrinsic property of polar active agents that extract their self-propulsion from a substrate and therefore align or anti-align their self-propulsion direction with their instantaneous velocity~\cite{Baconnier2025_rmp}. In liquid phases, this self alignment triggers collective motion~\cite{weber2013long,lam2015self}. In dense phases, it triggers collective actuation~\cite{Baconnier2022} as also found in living~\cite{Xu2023} and other biomimetic~\cite{Kellay2021, Levis2025, Ihle2026} systems. Applying external fields to such systems also reveals rich dynamical behaviors~\cite{Baconnier2025_prl}. In short, self-alignment recently opened the door to a broad phenomenology, which we leverage for unprecedented protocols of work extraction.

Our results combine experiments and numerical simulations of elastic solids made of self-aligning agents. We establish a phase diagram of work extraction, which demonstrates that active pistons extract work for a sufficiently slow compression and large activity. We also reveal the subtle role of fluctuations, which control the presence of a re-entrant transition. Once established experimentally and numerically that work extraction is not a many-body feature, we combine nonequilibrium response theory with heuristic arguments to derive the minimal necessary condition for an active piston to extract work. As a direct consequence, we show that the absence of coupling between the external perturbation and the internal degrees of freedom of active components precludes any work extraction when using active pistons, and how the specific relaxation mechanisms at play in self-alignment are liable to designing innovative engines. We elucidate the main mechanisms underlying engine performance, and provide physical intuition on how the competition between noise and activity regulates work extraction.


\begin{figure*}
	\centering
	\includegraphics[width=\textwidth]{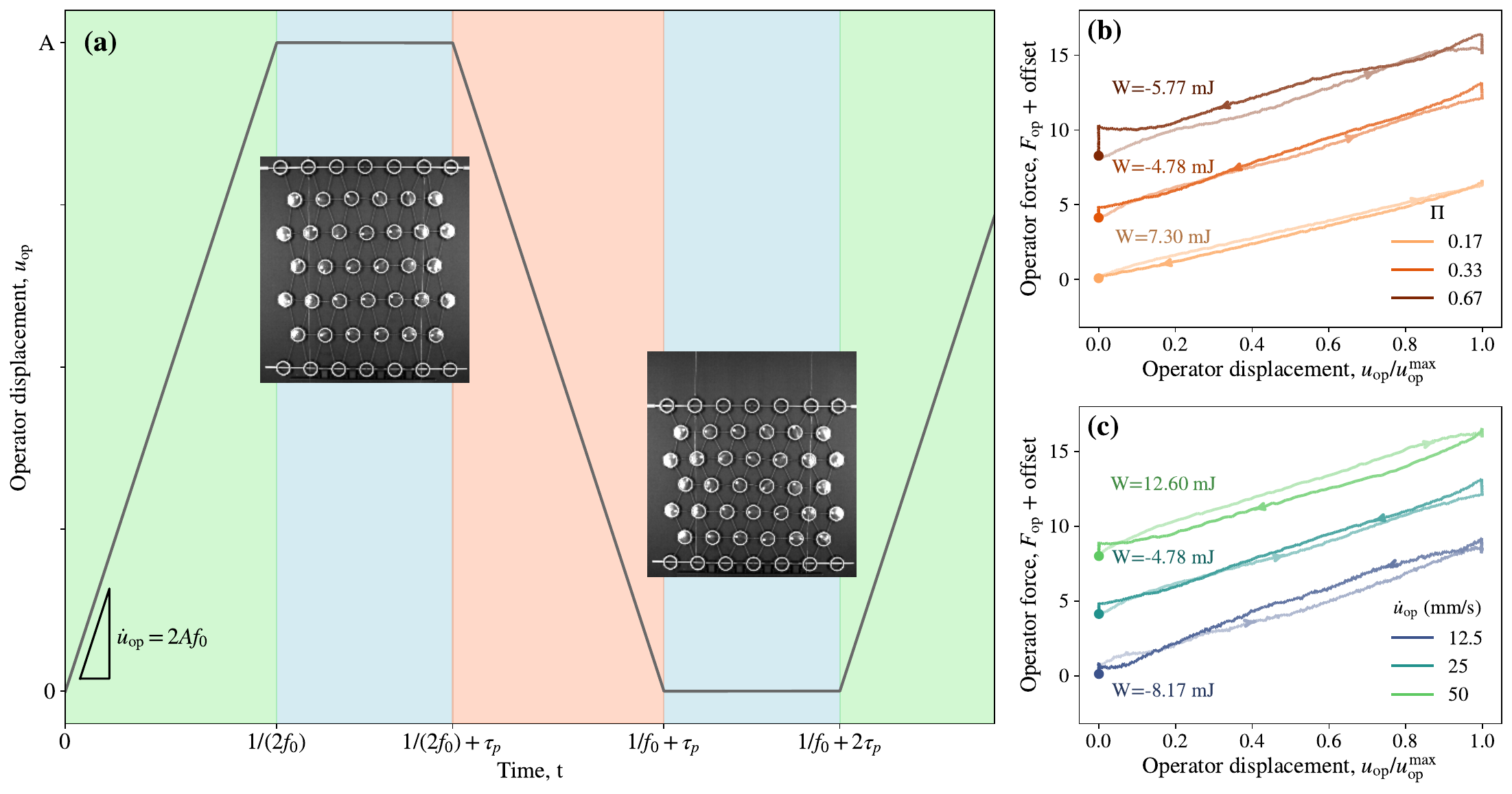}
	\caption{Experimental realization of active pistons.
	(a)~Periodic compression and expansion of an elastic network connecting hexbugs. The bottom layer is pinned, and the top layer is displaced by the operator. The forcing profile is composed of expansions and compressions at constant speed interceded with rest periods.
	(b,c)~For periodic expansion and compression, mechanical cycles emerge in the space of the dimensionless operator force $F_{\rm op}$ and operator displacement $u_{\rm op}$. Clockwise and counter-clockwise cycles supply work to ($W>0$) and extract work from ($W<0$) pistons, respectively.
	Parameters in Ref.~\cite{SM}.
	}
	\label{fig1}
\end{figure*}

\subsection*{Emergent mechanical cycles in active pistons}

We consider an active solid composed of polar active particles connected by springs in a triangular lattice. The corresponding experimental set-up, where the active particles are hexbugs~\cite{dauchot2019dynamics, Coulais2023} encaged in 3d-printed cylinders was studied in Ref.~\cite{Baconnier2022} with the lattice nodes at the boundary being clamped. Self-alignment is characterized by a length $l_a$ that describes the relaxation of the particle orientation onto the direction of its displacement~\cite{Baconnier2025_rmp}. Activity and elasticity set the elongation length $l_e$ of a spring by the self-propulsion force. The physics of polar active solids is thus controlled by the ratio of these two lengths, that is encapsulated in the activity parameter $\Pi=l_e/l_a$, and the amplitude of the active noise $D$~\cite{Baconnier2022}. Here, we design {\em active pistons} by clamping the bottom layer and imposing displacement of the top layer, so that all intermediate nodes are free to move [Fig.~\ref{fig1}(a)].

We perform periodic expansion [amplitude $A$, frequency $f_0$, speed $v_0=2 A f_0$, Fig.~\ref{fig1}(a)] and measure the force $F_{\rm op}(t)$ exerted by the operator when imposing a displacement $u_{\rm op}(t)$. When deforming the solid, self-alignment re-orients the self-propulsion forces towards the direction of driving ${\bf u}_{{\rm op}}$, which in turn reduces $F_{\rm op}$. The precise response of the solid depends on the organization of the self-aligning forces, which depends on the past history of the strain. Thus, one expects strongly nonlinear response, memory effects, and hysteresis, as reported in Figs.~\ref{fig1}(b,c), where the loading curves describe complex cycles in mechanical space $(F_{\rm op}, u_{\rm op})$.

\begin{figure*}
	\centering
	\includegraphics[width=\textwidth]{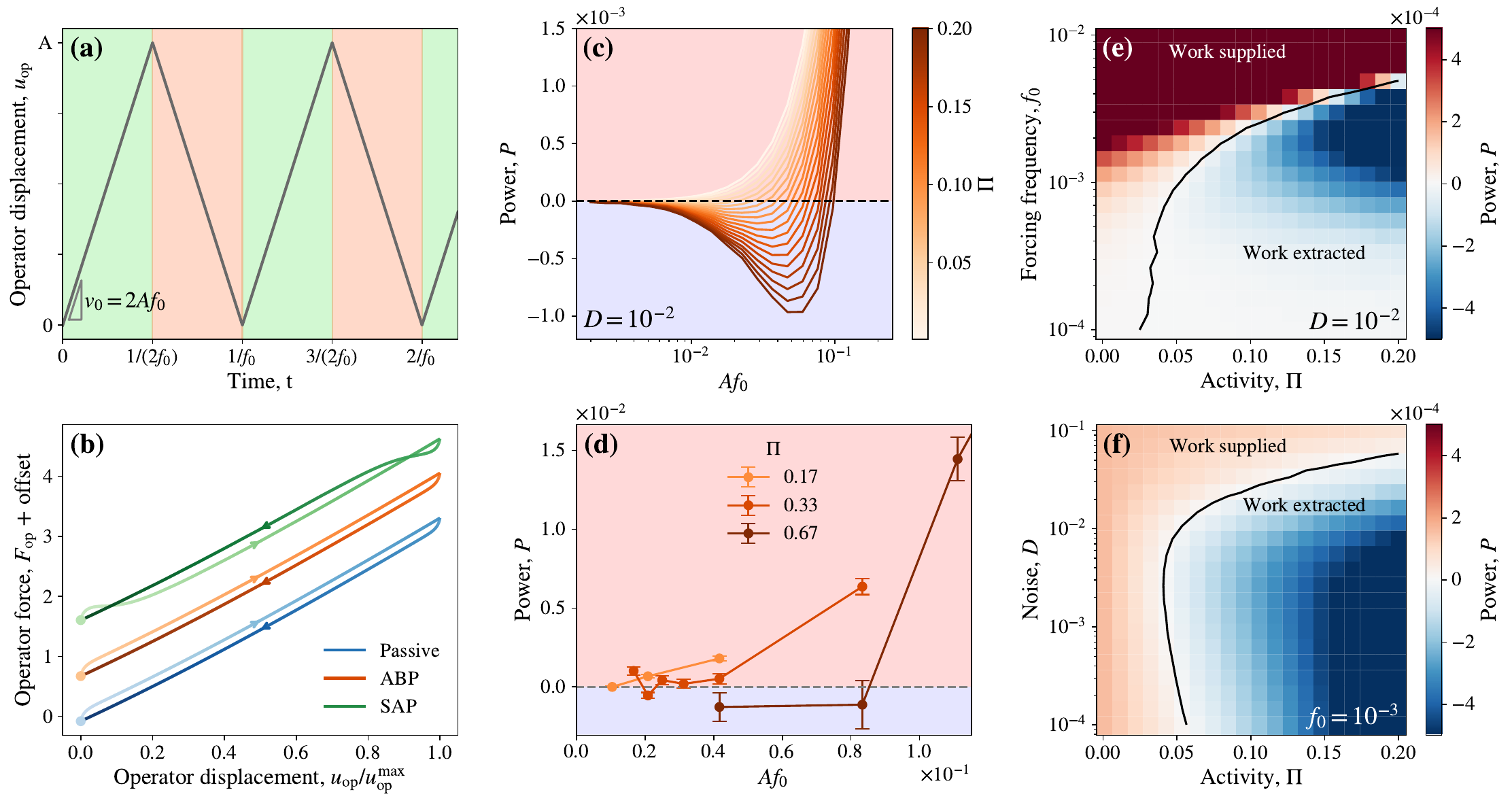}
	\caption{Theoretical model of two-dimensional active pistons.
	(a)~Periodic compression and expansion of an elastic network composed of motile self-aligning particles.
	(b)~For similar parameter values, a counter-clockwise mechanical cycle ($W<0$) for self-aligning particles [green, SAP] becomes clockwise ($W>0$) for motile particles without self-alignment [orange, ABP], and for passive particles without motility [blue, Passive].
	(c)~The operator power $P$ is negative at low frequency $f_0$ for high activity $\Pi$, in line with experimental results shown in panel (d).
	(e,f)~The phase diagram distinguishes regimes where work is either supplied to ($P>0$) or extracted from ($P<0$) active pistons. The non-monotonic dependence of $P$ with noise amplitude $D$, at fixed activity $\Pi$ and frequency $f_0$, yields a re-entrant transition.
	Parameter details in Ref.~\cite{SM}.
	}
	\label{fig2}
\end{figure*}

The work $W$ associated with these protocols coincides with the area of mechanical cycles:
\begin{equation}\label{eq:work}
	W = \int F_{\rm op} \, d u_{\rm op} .
\end{equation}
Work is supplied to the active piston ($W>0$) for clockwise cycles, and extracted ($W<0$) for counter-clockwise cycles. The cycle direction is an emergent property: in contrast with cyclic engines~\cite{Cates2021}, the operator only varies $u_{\rm op}$ without controlling any other parameters. Our experiments reveal that the shape and overall direction of the cycle change with compression speed $v_0$ and activity $\Pi$. For small $v_0$ and large $\Pi$, the active piston operates as an engine ($W<0$) [Figs.~\ref{fig1}(b,c)].


\subsection*{Self-alignment regulates work extraction}

We numerically study the mechanical properties of active pistons using the agent-based, overdamped, dimensionless dynamical equations introduced in Ref.~\cite{Baconnier2022} with proper boundary conditions. For a solid composed of $N+1$ layers, the location ${\bf r}_i$ and the orientation $\hat{\bf n}_i = (\cos\theta_i, \sin\theta_i)$ of the self-propulsion force at the nodes $i$ of the $N-1$ free layers obey~\cite{SM}
\begin{equation}\label{eq:dyn}
\begin{aligned}
	\dot{\bf r}_i &= \Pi \, \hat{{\bf n}}_i + \sum_{j\in\partial i} F_{ij} \hat{{\bf e}}_{ij},
	\\
	\dot \theta_i &= (\hat{\bf n}_i \times \dot{\bf r}_i )\cdot \hat{\bf e}_z+ \sqrt{2D} \, \eta_i ,
\end{aligned}
\end{equation}
where $F_{ij}$ is the amplitude of the elastic force between two nodes connected along the unit vector $\hat{{\bf e}}_{ij}$. The bottom layer is fixed, while the top layer is driven by the operator~\cite{SM}. The Gaussian white noise $\eta_i$ has zero mean and unit variance, while the noise amplitude $D$ controls the orientation relaxation. The self-alignment mechanism is embodied by the term $(\hat{{\bf n}}_i \times \dot{\bf r}_i) \cdot\hat{{\bf e}}_z$.

Under periodic compression [Fig.~\ref{fig2}(a)], the emergent mechanical cycles $(F_{\rm op}, u_{\rm op})$ obtained for passive systems [$\Pi=0$] and for active systems without self-alignment [$\Pi\neq 0$ and $\dot{\theta}_i = \sqrt{2D}\, {\eta}_i$ in Eq.~\eqref{eq:dyn}] are identical: they always operate clockwise ($W>0$), showing that self-propulsion without self-alignment is not sufficient for work extraction [Fig.~\ref{fig2}(b)]. In other words, our active pistons specifically leverage the mechanisms at play in self-alignment.

The dependence of the extracted power $P = W f_0$ on the cycling frequency $f_0$ [Fig.~\ref{fig2}(c)] reveals that active pistons extract work for large enough activity and small enough $f_0$ . Experiments confirm this trend [Fig.~\ref{fig2}(d)] in line with the shape changes of mechanical cycles reported on [Figs.~\ref{fig1}(c,d)]. Figs.~\ref{fig2}(e,f) report the phase diagrams that systematically delineate the frontier for work extraction in terms of $(f_0, \Pi, D)$: we observe a re-entrance indicating that work is extracted only at intermediate values of $D$.

\begin{figure*}
	\centering
	\includegraphics[width=\textwidth]{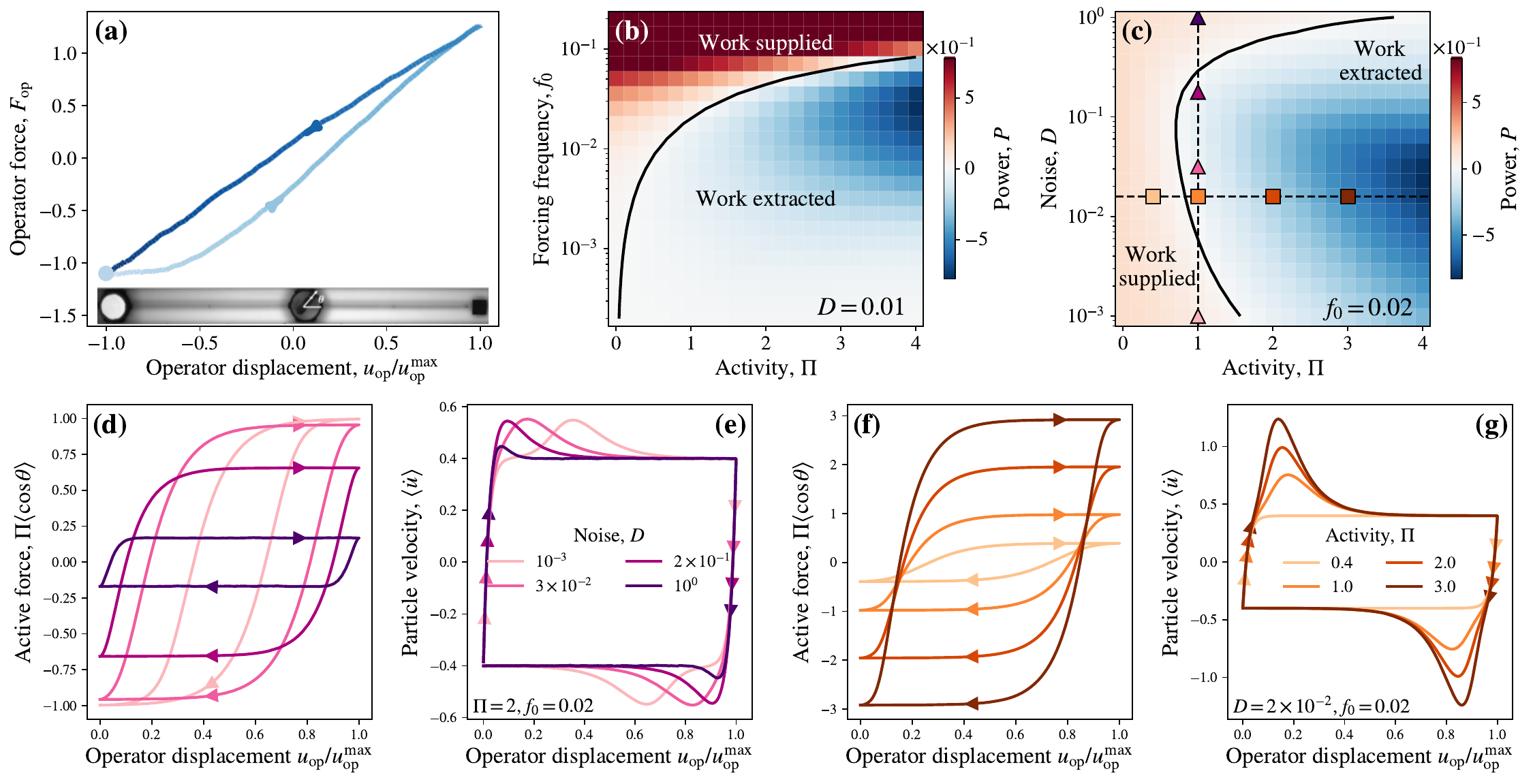}
	\caption{One-dimensional active pistons with a single hexbug.
	(a)~Experiments reveal a counter-clockwise mechanical cycle showing work extraction $W<0$.
	(b,c)~Phase diagrams in theoretical models reproduce the same qualitatively features as in two-dimensional active pistons with many hexbugs [Figs.~\ref{fig2}(e,f)], including a re-entrant transition with the noise amplitude $D$. Markers refer to mechanical cycles in panels~(d-g).
	The active force $\Pi\langle\cos\theta\rangle$ and the particle velocity $\langle\dot u\rangle$ always feature clockwise cycles as functions of the operator displacement $u_{\rm op}$, whose complex dependence on noise amplitude $D$ and activity $\Pi$ helps rationalize the re-entrant transition in the phase diagram.
	Parameter details in Ref.~\cite{SM}.
	}
	\label{fig3}
\end{figure*}

In view of theoretical analysis, a matter of interest is whether collective effects are at play in the mechanism for work extraction. It is actually not the case: the experimental loading curve obtained for a single hexbug confined in a rail and connected to two springs, one clamped and the other driven by the operator, also features counter-clockwise mechanical cycles [Fig.~\ref{fig3}(a)], showing that work can be extracted from periodic compression as in the two-dimensional version [Figs.~\ref{fig1}(c,d)].

The dynamical equation in this setting drastically simplifies, with only two degrees of freedom, namely the displacement $u$ and the
orientation $\theta$ of the hexbug, without any geometric nonlinearities:
\begin{equation}\label{eq:rail}
\begin{aligned}
	\dot u &= \Pi \cos \theta - 2 u + u_{\rm op} ,
	\\
	\dot \theta & = (h - \dot u) \sin \theta + \sqrt{2D} \, \eta .
\end{aligned}
\end{equation}
The term $h$ describes a polarizing field applied to the orientation~\cite{Baconnier2025_prl, Baconnier2025_pre}, which we shall use below when developing linear response theory. Performing numerical simulations of these equations (at $h=0$), we recover the same phase diagrams as for the full 2D solid [Figs.~\ref{fig3}(b,c)]. In short, experiments and simulations demonstrate that the regimes of work extraction are qualitatively identical for either two-dimensional pistons with many hexbugs or one-dimensional pistons with a single hexbugs. We now leverage this equivalence to delineate the minimal ingredients for work extraction.

\begin{figure*}
	\centering
	\includegraphics[width=\linewidth]{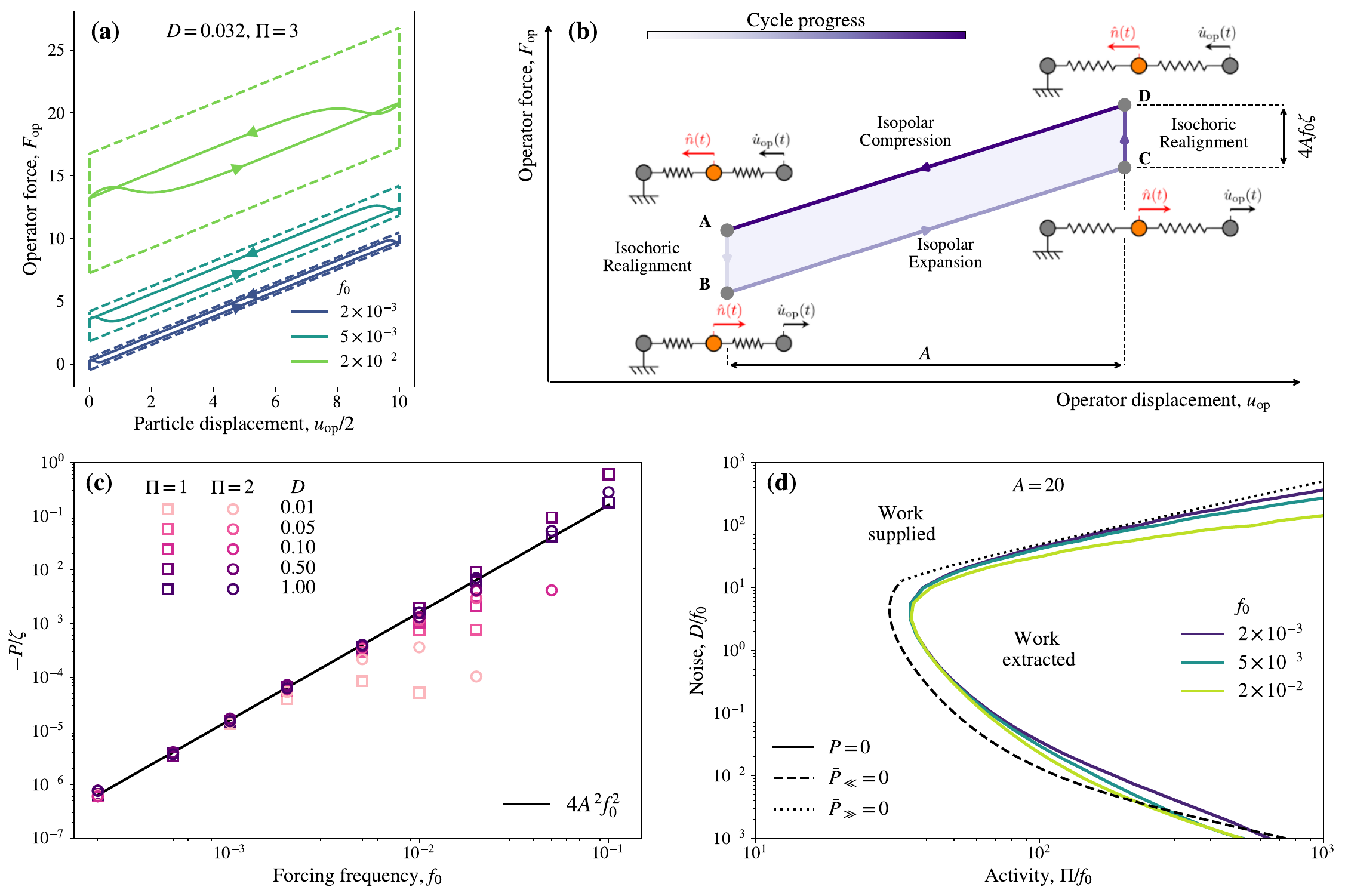}
	\caption{Insights from linear response theory.
	(a)~Mechanical cycles in numerical simulations (solid lines) converge at low frequency $f_0$ to the parallelogram predicted by linear response (dashed lines).
	(b)~During compression and expansion (B$\to$C and D$\to$A), the self-propulsion orientation $\hat n$ points along the displacement direction $\hat u_{\rm op}$: it yields affine relations between the operator force $F_{\rm op}$ and displacement $u_{\rm op}$. When switching between compression and expansion (A$\to$B and C$\to$D), $\hat n$ relaxes (almost) immediately towards $\hat u_{\rm op}$ at constant displacement. The cycle area is determined by the amplitudes of displacement $A$ and of realignment $A f_0 \zeta$, where $\zeta$ is the linear response coefficient.
	(c)~The extracted power $P$ scaled by $\zeta$ follows the master curve $4(Af_0)^2$ at low $f_0$ independently of material properties $(\Pi,D)$ within linear response.
	(d)~The phase boundary of work extraction ($P=0$, numerical simulations) follows a master curve $(D/f_0, \Pi/f_0)$ qualitatively captured by combining linear response ($\bar P_\gg=0$) and heuristic ($\bar P_{\ll}=0$) derivations; see Ref.~\cite{SM} for other values of $A$.
	Parameter details in Ref.~\cite{SM}.
	}
	\label{fig4}
\end{figure*}

The work [Eq.~\eqref{eq:work}] can be simplified as $W = W_{\rm fr} + W_{\rm ac}$, where the friction and active terms read
\begin{equation}\label{eq:decomp}
	W_{\rm fr} = \frac12 \int \langle \dot u \rangle \, d u_{\rm op} ,
	\quad
	W_{\rm ac} = - \frac\Pi2 \int \langle \cos\theta \rangle \, d u_{\rm op} ,
\end{equation}
and $\langle\cdot\rangle$ denotes an average over noise realizations. In fact, the elastic contribution to the work vanishes over a cycle~\cite{SM}. The transition between extracted and supplied work follows by examining the shape of emergent cycles in the appropriate mechanical space: $(\langle\dot u\rangle, u_{\rm op})$ for $W_{\rm fr}$, and $(\langle\cos\theta\rangle, u_{\rm op})$ for $W_{\rm ac}$. The corresponding cycles are always clockwise, so the active and friction terms respectively extract ($W_{\rm ac}<0$) and supply ($W_{\rm fr}>0$) work [Figs.~\ref{fig3}(d-g)]. Therefore, delineating regimes of work extraction amounts to comparing the cycle areas that regulate the competition between $W_{\rm ac}$ and $W_{\rm fr}$.

At vanishing noise ($D=0$), the orientation depends on $u_{\rm op}$ through a univocal relation $\theta=\theta(u_{\rm op})$~\cite{SM}. Accordingly, the compression and expansion in $(\langle\cos\theta\rangle, u_{\rm op})$ are undistinguishable, so $W_{\rm ac}$ vanishes and pistons never extract work. Increasing $D$ speeds up the orientation relaxation, so the compression and expansion branches now differ. In fact, cycles in both $(\langle\dot u\rangle, u_{\rm op})$ and $(\langle\cos\theta\rangle, u_{\rm op})$ converge towards a rectangular shape at large $D$ [Figs.~\ref{fig3}(d-e)], which reflects that pistons relax (almost) instantaneously as $u_{\rm op}$ switches between compression and expansion. The horizontal lines in these cycles correspond to stationary values: $\langle\cos\theta\rangle$ decreases with $D$ in steady-state~\cite{Baconnier2024}, while $\langle\dot u\rangle$ converges to $\dot u_{\rm op}/2$ for all $D$. Therefore, the cycle area in $(\langle\cos\theta\rangle, u_{\rm op})$ decreases at small and large $D$, whereas the one in $(\langle\dot u\rangle, u_{\rm op})$ saturates at large $D$. In all, these behaviors explain the re-entrance in the phase diagram: $W_{\rm fr}$ overwhelms $W_{\rm ac}$ at small and large $D$, so pistons only extract work at intermediate $D$  [Fig.~\ref{fig3}(c)].

At large $\Pi$, the cycle area in $(\langle\cos\theta\rangle, u_{\rm op})$ increases faster than in $(\langle\dot u\rangle, u_{\rm op})$ [Figs.~\ref{fig3}(f-g)], so pistons always extract work above a critical activity. Here, the drive $u_{\rm op}$ affects the orientation $\theta$ due to self-alignment; in other types of active particles~\cite{Marchetti2018}, the absence of self-alignment decouples $\theta$ from $u_{\rm op}$, which yields a vanishing $W_{\rm ac}$ and precludes work extraction. From a broader perspective, we expect that any active system that couples displacement (and thus compression) with some internal degrees of freedom, either orientation~\cite{Baconnier2025_rmp} or other configuration variables~\cite{Golestanian2021, Zhang2023, Cocconi2025}, is amenable to work extraction through periodic compression alone.

In what follows, we show that a linear response approach for slow protocols captures the essence of the mechanism underlying work extraction. Beyond linear response, heuristic arguments allow us to understand the re-entrant transition in terms of noise amplitude.


\subsection*{Competition between noise and activity}

Active pistons always extract work at high activity for low enough frequency. To rationalize the mechanism regulating the competition between noise and activity in the phase diagram, we here propose a perturbative approach at very low frequency. In this regime, we reveal that mechanical cycles systematically approach a parallelogram shape [Fig.~\ref{fig4}(a)] characterized by only three independent parameters: the cycle frequency and amplitude, and a material coefficient. Specifically, we demonstrate that nonequilibrium linear response~\cite{Baiesi2013, Davis2024} leads to identifying the relevant material coefficient, and helps rationalize why the displacement-orientation coupling, inherent to self-alignment, is key to operating our pistons.

Assuming that the piston operates slowly, we express the force $F_{\rm op}$ within linear response as
\begin{equation}\label{eq:force}
	F_{\rm op} \approx u_{\rm op}/2 + \zeta\, \dot u_{\rm op} ,
\end{equation}
in terms of the material coefficient $\zeta = \int_0^\infty \frac{\delta\langle u\rangle(t)}{\delta u_{\rm op}(0)} \,t \,dt$. This relation holds for an arbitrary piston independently of the underlying dynamics. For our specific case [Eq.~\eqref{eq:rail}], we demonstrate that $\zeta$ reads~\cite{SM}
\begin{equation}\label{eq:zeta}
	\zeta = \frac14 \bigg( 1 - \Pi \, \frac{\partial\langle\cos\theta\rangle}{\partial h} \bigg) ,
\end{equation}
where the static susceptibility $\frac{\partial\langle\cos\theta\rangle}{\partial h}$ is evaluated in the unperturbed dynamics ($\dot u_{\rm op}=0$) for a small polarizing field ($h\to0$). This susceptibility describes how $F_{\rm op}$ is actually regulated by the orientation $\theta$. Without self-alignment [$\Pi\neq0$ and $\dot\theta = \sqrt{2D} \eta$ in Eq.~\eqref{eq:rail}], the material coefficient reduces to $\zeta=1/4$~\cite{SM}: the absence of displacement-orientation coupling precludes any non-trivial dependence of linear response on material properties. In short, $\zeta$ encapsulates the competition between mechanical friction and self-alignment that determines whether work is supplied or extracted.

For quasistatic protocols in which pistons evolve slower than any relaxation timescale, there is a linear relation between force and displacement: $F_{\rm op} \approx u_{\rm op}/2$ since $\dot u_{\rm op}\approx 0$. Consequently, the loading curves are identical for compression and expansion, so mechanical cycles are reduced to a single line without work extraction. Instead, protocols operating at a finite constant (slow) rate $\dot u_{\rm op} = \pm 2 A f_0$ yield an affine relation between force and displacement [Eq.~\eqref{eq:force}]. In this regime, mechanical cycles consist of four branches: (i)~two oblique lines when compression and expansion change displacement $u_{\rm op}$ at a constant orientation $\theta$, and (ii)~two vertical lines for change in orientation $\theta$ at a constant $u_{\rm op}$ [Fig.~\ref{fig4}(b)]. Changes in $\theta$ operate much faster than changes in $u_{\rm op}$ when the protocol switches between compression and expansion.

We now discuss how the shape of mechanical cycles affects work extraction within linear response. The force expansion [Eq.~\eqref{eq:force}] is consistent with approximating $\langle \dot u \rangle$ by $\dot u_{\rm op}/2$; in this regime, the work decomposition [Eq.~\eqref{eq:decomp}] simplifies as $W_{\rm fr} = A^2 f_0 + {\cal O}(f_0^2)$ and $W_{\rm ac} = (4\zeta-1) A^2 f_0 + {\cal O}(f_0^2)$~\cite{SM}. Therefore, $W = W_{\rm fr} + W_{\rm ac}$ is entirely determined by the cycle parameters $(A,f_0)$ and the linear response coefficient $\zeta$ to leading order in $f_0$; indeed, these are the only parameters that regulate the cycle area in the space $(F_{\rm op}, u_{\rm op})$ [Fig.~\ref{fig4}(a)]. The extracted power $P = W f_0$ follows as $P = 4 \zeta (A f_0)^2 + {\cal O}(f_0^3)$. Our numerical simulations confirm that $P/\zeta$ falls into a master curve independent of material properties $(D,\Pi)$ at low $f_0$ [Fig.~\ref{fig4}(c)].

To rationalize how noise regulates work extraction beyond linear response, we search for an analytical prediction of the form
\begin{equation}\label{eq:scaling}
	P = (Af_0)^2 \, \bar P(A, \Pi/f_0, D/f_0) .
\end{equation}
In the regime of small $\Pi$, we derive two approximate solutions: $\bar P = \bar P_\ll (A,\Pi/f_0,D/f_0)$ for small $D$, and $\bar P = \bar P_\gg (A,\Pi/f_0,D/f_0)$ for large $D$~\cite{SM}. The phase boundary for work extraction follows by matching the conditions of vanishing power from $\bar P_\ll=0$ and $\bar P_\gg=0$; we set the regime of validity for each approximation by enforcing that $(\bar P_\ll, \bar P_\gg)$ have the same sign. This procedure reveals a master curve in the phase diagram $(D/f_0, \Pi/f_0)$ that is qualitatively comparable with our numerics [Fig.~\ref{fig4}(d)]. Note that expanding $\bar P_\gg$ at large $D/f_0$ recovers linear response $P = 4 \zeta (A f_0)^2$ with $4\zeta = 1 - \Pi/(2 D)$, yet it is inadequate to capture any re-entrance in the phase diagram~\cite{SM}.

In short, combining two distinct predictions that are both consistent with small $\Pi$, we put forward a master curve that describes the re-entrance of the phase diagram in line with numerical results. This curve shows that, at every frequency, active pistons extract work only when the activity exceeds a threshold set by the noise; below this threshold, the noise overwhelms the self-alignment, so the activity mechanism effectively reduces to self-propulsion only, which precludes work extraction.


\subsection*{Discussion}

Active pistons embody a novel class of nonequilibrium engines, without any equivalent in thermal systems, operating solely through volume control. In contrast to standard active engines that extract work from cyclic protocols, our active pistons do not require changing any bulk properties, since control of volume through boundaries here suffices for work extraction: this convenient feature opens a wide range of opportunities for engine design far from equilibrium.

Our results help delineate the minimal conditions for pistons to extract work. The key ingredient is that self-propulsion orientation couples to compression. Here, we have shed light on self-alignment as a useful design principle, and shown that the regime of work extraction extends towards low frequency at high activity. From a broader perspective, linear response theory offers guidelines for building other active pistons: we have shown that measuring the appropriate response coefficient leads to identifying specific active systems prone to extract work from periodic compression alone. Along this line, we anticipate that the negative mobility of tracers embedded in active baths~\cite{Granek2022, Baek2024, Maes2026} can be usefully leveraged, while it remains to determine whether odd materials~\cite{Souslov2021, Fruchart2023, Fakhri2026} are amenable to design pistons similar to ours.

The master curve of the phase diagram with noise re-entrance lies beyond linear regimes, so it remains to explore further whether it is a generic feature beyond our active pistons. Our results call for future studies to examine protocol optimization using recent developments in nonequilibrium control theory~\cite{Olsen2025, Loos2025, Alvarado2026}: how to optimize finite-time compression to increase work extraction. In this context, linear response theory provides the groundwork for a systematic study~\cite{Davis2024, Soriani2025, Zhong2026}. Finally, our pistons can inspire the design of novel information engines far from equilibrium~\cite{Saha2023, Leighton2025, Goerlich2025}.

This project has received funding from the Luxembourg National Research Fund (FNR) grant references 14389168 and 18118949, and from the French National Research Agency (ANR) grant reference ANR-24-CE30-6841-01. \'E.F. and O.D. acknowledge support from Grant No. NSF PHY-2309135 to the Kavli Institute for Theoretical Physics (KITP). E.K. acknowledges individual KIAS Grant No.~QP10301 at the Korea Institute for Advanced Study. P.B. acknowledges ED564 "Physique Ile de France" for their PhD grant.


\bibliography{references}


\clearpage
\onecolumngrid

\renewcommand{\theequation}{S\arabic{equation}} 
\renewcommand{\thesection}{S\arabic{section}}  
\renewcommand{\thetable}{S\arabic{table}}  
\renewcommand{\thefigure}{S\arabic{figure}}	
\setcounter{equation}{0}
\setcounter{section}{0}
\setcounter{figure}{0}
\setcounter{table}{0}

\begin{center}
{\large\bf Supplementary materials: Active pistons extract work by periodic compression alone}
\\[10pt]
Paul Bernard,$^{1}$ Euijoon Kwon,$^{2}$ Benjamin Buhl,$^{1,3}$ \'Etienne Fodor,$^{4}$ and Olivier Dauchot$^{1}$
\\[4pt]

\textit{$^{1}$Gulliver UMR CNRS 7083, ESPCI Paris, PSL Research University, 10 rue Vauquelin, 75005 Paris, France}
\\
\textit{$^{2}$Quantum Universe Center, Korea Institute for Advanced Study, Seoul 02455, Republic of Korea}
\\
\textit{$^{3}$Institut Lumière Matière UMR5306 - UCBL - CNRS, 10 rue Ada Byron, 69622 Villeurbanne, France}
\\
\textit{$^{4}$Department of Physics and Materials Science, University of Luxembourg, L-1511 Luxembourg City, Luxembourg}

\end{center}
\vspace{1em}



\section{Two-dimensional active pistons}

In this Section, we describe the experimental and numerical methods used to obtain the results presented in the main for the active pistons made of many hexbugs embedded in a two-dimensional elastic lattice.


\subsection{Lattice experiments}
\subsubsection{Setup description}

\begin{figure}[b]
    \centering
    \includegraphics[width=0.6\linewidth]{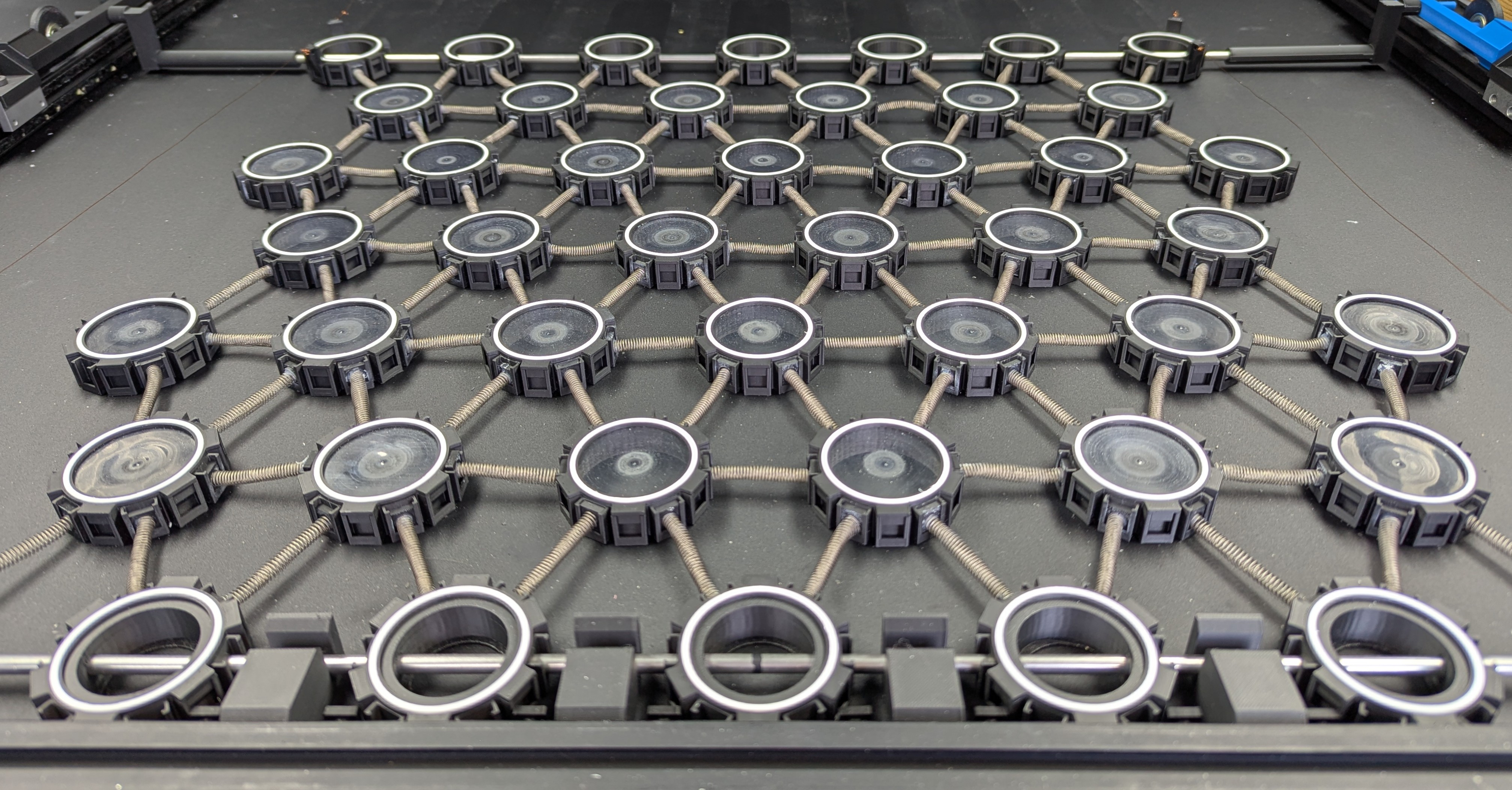}
    \caption{Two-dimensional elastic lattice constructed using Hookean Springs and 3D-printed annuli that contain hexbugs. The top and bottom layers do not house hexbugs. The bottom layer is pinned. The top layer is skewered with a steel rod, and its displacement is controlled by the operator using cables that are wound and unwound with stepper motors.}
    \label{fig:exp_lattice}
\end{figure}

Polar active lattices are constructed using 3D-printed annuli, that are interconnected using Hookean Springs. The Hookean springs are purchased from Schweizer FerernTechnik, where wands have rest length $l_0=4$~cm and stiffness $k=2.4\,\text{N/m}$. To make stiffer bonds, the springs can be intertwined to effectively act in parallel, and leverage the stiffness additivity of parallel springs. Double and quadruple springs are so made, with respective stiffnesses $4.8\,\text{N/m}$ and $9.6\,\text{N/m}$ for the same rest length. Annuli are assembled into a triangular lattice structure [Fig.~\ref{fig:exp_lattice}]. The observation arena is enclosed by a rigid frame. The bottom layer of the lattice is kept steady by attaching it to the frame. The top layer is driven by the operator using metal cables, that are wound and unwound using pulleys controlled by stepper motors. Stepper motors are piloted using Arduino and A4988 microstep drivers.

Hexbugs are placed within the annuli in the bulk, forming the elementary active agent. Hexbug characterization has been detailed in Ref.~\cite{Baconnier2022}. The hexbug loaded annulus follows overdamped dynamics (inertial time of $0.12\pm0.01\,s$), that is characterized by three parameters: the active force $F_a = 43\pm3\,\text{mN}$, the alignment length $l_a = 2.5\pm0.3\,\text{cm}$, and the angular diffusion $D=1.75\pm0.15\,\text{rad}^2/\text{s}$. Using the hexbug and lattice parameters, we can obtain the activity parameter $\Pi=F_a/(kl_a)$. In our experiments, the activity parameter is modified by changing the springs' stiffness. The studied lattices have $N=32$ active nodes. Forcing cycles are constitutedmade constant speed tractions and compressions with pause phases [Fig.~1 in main text]. The amplitude of forcing is of $A=15$~cm for spring stiffness of $2.4\,\text{N/m}$ and $4.8\,\text{N/m}$, and $A=10$~cm for stiffness $9.6\,\text{N/m}$ (due to torque constraints on the used motors).


\subsubsection{Force and work measurement}
 
\begin{figure}
	\centering
	\includegraphics[width=\linewidth]{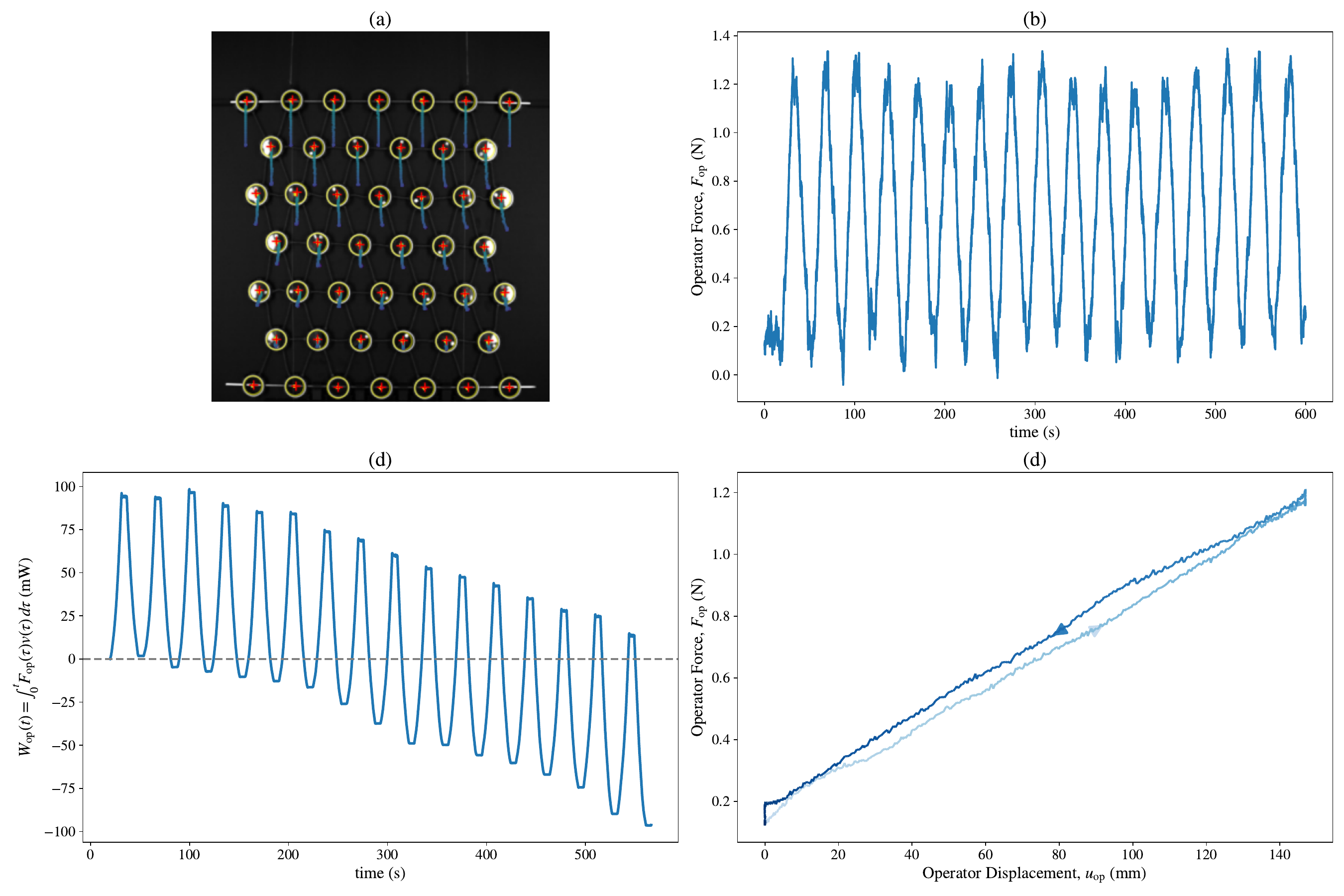}
	\caption{Experimental measurement of the work produced by the external operator.
	(a)~Feature detection in the polar active lattice where the lower layer is pinned, and the top layer is moved by the operator.
	(b)~Operator force $F_{\rm op}$ measured over time by tracking spring elongations and summing all the elastic forces applied to the operator nodes.
	(c)~Cumulative work $W_{\rm op}$ applied by the operator to the active lattice over an experiment of $10$~min. The cumulative work becomes negative for many cycles, signifying that the operator extract work from the active solid.
	(d)~Corresponding average mechanical cycle $(F_{\rm op}, u_{\rm op})$. Parameters: $\Pi=0.33$, $f_0 = 4.2\times 10^{-2}$~Hz, and $A=15$~cm.
    }
	\label{fig:work_meas}
\end{figure}

The nodes of the lattice are tracked by tracking the white circular inset [Fig.~\ref{fig:exp_lattice}]. The circular insets are detected using the Hough Transform of the OpenCV library in Python, with a positional error of $0.5$~mm. The node positions are then connected together using the Crocker-Grier algorithm~\cite{CROCKER1996298}. Tracking the positions in time, it is then possible to calculate the elastic forces applied to each node from the classical Hooke spring force: ${\bf F}_{ij} = F_{ij} \hat{\bf e}_{ij}$, where $F_{ij} = k(r_{ij}-l_0)$ is defined in terms of the internode distance $r_{ij}$, and $\hat{\bf e}_{ij}$ is a unit vector connecting the nodes $(i,j)$. The operator force $F_{\rm op}$ is defined as the total elastic force applied to the top layer controlled by the operator, from which follows the work $W_{\rm op}$ as
\begin{equation}\label{eq:observabels_exp}
	F_{\mathrm{op}} = \sum_{i\in \text{op}}\sum_{j\in \partial i}F_{ij} ,
	\qquad
	W_{\mathrm{op}} = \int_0^{1/f_0}dt F_{\mathrm{op}}(t)\dot u_{\mathrm{op}}(t) ,
\end{equation}
where $f_0$ is the cycle frequency, and $u_{\rm op}$ the displacement of the top layer. An example of the work measurement is presented in Fig.~\ref{fig:work_meas}. Work measurements are averaged over at least 40 measurements per set of parameters [Tab.~\ref{tab:tau_k_summary}].

\begin{table}[b]
\centering
\begin{tabular}{ccrr}
\toprule
$\Pi$ & $f_0$ & Cycles &  Experiments \\
\hline
\midrule
\multirow{4}{*}{0.67} & 0.1667  & 98  & 3 \\
                    & 0.1111  & 136 & 5 \\
                    & 0.0833  & 69  & 3 \\
                    & 0.0417  & 42  & 3 \\
\hline
\midrule
\multirow{6}{*}{0.33} & 0.1667  & 140 & 4 \\
                    & 0.0833  & 119 & 5 \\
                    & 0.0625  & 96  & 5 \\
                    & 0.0500  & 75  & 5 \\
                    & 0.0417  & 55  & 4 \\
                    & 0.0333  & 45  & 4 \\
\hline
\midrule
\multirow{3}{*}{0.17} & 0.2500 & 117 & 3 \\
                    & 0.1250 & 59  & 2 \\
                    & 0.0625 & 76  & 4 \\
\bottomrule
\end{tabular}
\caption{Statistics of the experimental measurements for different values of $(\Pi,f_0)$. The third and fourth columns respectively correspond to the total number of forcing cycles over the respective number of 10-minute experimental runs used during measurement.}	
\label{tab:tau_k_summary}
\end{table}

\subsection{Lattice numerics}

\begin{figure}
    \centering
    \includegraphics[width=\linewidth]{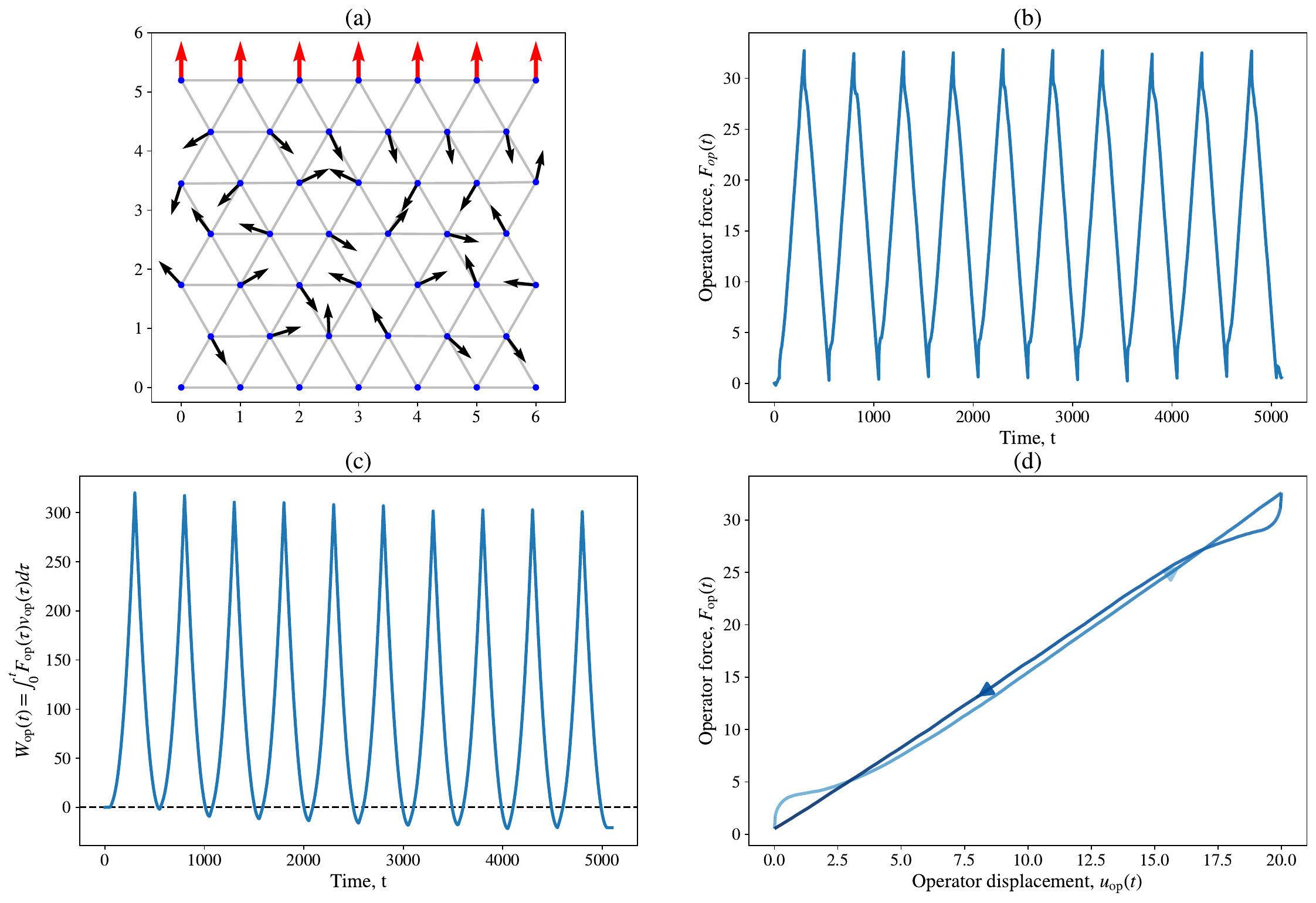}
    \caption{Measuring work in numerical simulations.
    (a)~Initialization of self-aligning polar lattice, with the same dimensions as the experimental lattice ($N=32$). The bottom layer is pinned, and the top layer displaced in the axial direction by the operator (red arrow). The bulk follows the positional and orientational dynamics. The orientations (black arrows) are initialized randomly according to a uniform distribution.
    (b)~Operator force $F_{\rm op}$ measured over time from spring elongations, and summing all the elastic forces applied to all the operator nodes.
    (c)~Cumulative work $W_{\rm op}$ applied by the operator to the active lattice over an experiment of $10$~min.
    (d)~Corresponding average mechanical cycle $(F_{\rm op},u_{\rm op})$. Parameters (in simulation units): $l_e = 0.015$, $l_a=0.1$, $D=0.01$, $A=2.0$, $f_0 = 2\times10^{-3}$.}
    \label{fig:numerical_lattice}
\end{figure}

In numerical simulations, active lattices are generated in a similar way as in the experimental lattice: nodes are decomposed into layers, where the bottom layer is pinned, and the top layer is displaced by the operator, whereas the bulk layers follow a self-aligning active dynamics [Fig.~\ref{fig:numerical_lattice}]. Specifically, the operator and pinned nodes obey
\begin{equation}
	u_i (t) =
	\begin{cases}
		u_{\mathrm{op}}(t) \quad\text{if}\quad i\in\text{operator nodes} ,
		\\
		0 \quad\qquad\text{if}\quad i\in\text{pinned nodes} .
\end{cases}
\end{equation}
The overdamped dynamics for the positions ${\bf r}_i$ and orientations $\hat{\bf n}_i = (\cos\theta_i, \sin\theta_i)$ of the nodes in the bulk layers is given by
\begin{equation}\label{eq:dynamics}
	\gamma \dot{\bf r}_i = F_a \hat{\bf n}_i + \sum_{j\in \partial i} k(r_{ij} - l_0) \hat{\bf e}_{ij} ,
	\quad
	\dot \theta_i = \dfrac{\alpha}{l_a} (\hat{\bf n}_i \times \dot{\bf r}_i)\cdot \hat{\bf e}_z + \sqrt{2D} \eta_i ,
\end{equation}
where $\eta_i$ is a Gaussian white noise with unit variance and zero mean. We take $\alpha=0$ and $\alpha=1$ for active Brownian particles (i.e., ABPs without self-alignment~\cite{Marchetti2018}) and for self-aligning particles (i.e., SAPs~\cite{Baconnier2022}), respectively. Lengths are rescaled by the rest length of springs $l_0$, and times by the damping timescale $\gamma/k$, yielding the following dimensionless dynamics
\begin{equation}\label{eq:dimensionless_dynamics}
	\dot{\bf r}_i = l_e \hat{\bf n}_i + \sum_{j\in \partial i} (r_{ij} - 1) \hat{\bf e}_{ij} ,
	\quad
	\dot \theta_i = \dfrac{\alpha}{l_a} (\hat{\bf n}_i \times \dot{\bf r}_i )\cdot \hat{\bf e}_z+ \sqrt{2D} \eta_i .
\end{equation}
This dynamics is simulated using a first order Euler-Maruyama scheme. Contrary to prior works on polar active solids~\cite{Baconnier2025_rmp}, the harmonic regime is not enforced in the simulation, as we expect the reference positions to change as the operator acts on the lattice. Operator force and work are defined as in the experimental case [Eq.~\eqref{eq:observabels_exp}]. In the main text, $l_a$ is chosen as the unit length, for consistency with~\cite{Baconnier2022}. Numerical results shown for polar active lattices in Fig.~2 of the main text correspond to parameter values reported in Tab.~\ref{tab:Lattice_sim_params}.

\begin{table}[b]
	\centering
	\begin{tabular}{c|c|c|c|c|c}
	& $l_e$ & $l_a$ & $D$ & $A$ & $f_0$
	\\
    \hline
	Fig.~2(b) Passive/SAP/ABP & $0/0.015/0.015$ & x$/0.1/$x & x$/0.01/0.01$  & $2.0$ & $2.0 \times 10^{-3}$\\
    \hline
	Fig.~2(c), 2(e) & $[0,0.02]$ & $0.1$ & $0.01$ & $2.0$ & $[10^{-4},10^{-2}]$
	\\
	\hline
	Fig.~2(f) & $1.5\times10^{-2}$ & $0.1$ & $[10^{-4},10^{-1}]$ & 2.0 & $10^{-3}$
	\\
	\end{tabular}
	\caption{Simulation parameters used for lattice results in Fig.~2 of the main text where lengths are rescaled by $l_a$.}
	\label{tab:Lattice_sim_params}
\end{table}


\newpage

\section{One-dimensional active pistons}

In this Section, we examine the performance of one-dimensional active pistons with a single hexbug. 

\subsection{Single particle experiments}\label{sec:exp}

The one-dimensional active piston is constructed with a hexbug embedded in a 3D-printed annulus as a model system of an active particle. The annulus is confined between two parallel metallic walls to make its displacement one-dimensional; see Fig.~3(a) in the main text. The annulus is fixed, thanks to two Hookean springs with stiffness $k = 1.2$~N/m and rest length $l_0 = 8$~cm, to a fixed wall on the left side and to a mobile structure on the opposite side. Springs are under tension during the entire experiment to avoid buckling. The mobile structure is moved by an non-extensible copper string attached to a pulley, turning forth and back by a stepper motor controlled with an ARDUINO UNO card. We force the mobile structure to follow a sinusoidal displacement $u_{\rm op} (t)=  A \sin \left(\frac{2 \pi}{T} t \right)$, with controlled temporal period $T = 6$~s and amplitude of $A = 9$~cm. Considering the spring stiffness and the hexbug's properties discussed previously, the active parameter in this experiment is $\Pi = 0.7$.


\subsection{Stochastic dynamics and deterministic regime}

Setting $l_a$ as the unit length and $\gamma/k$ as the unit time, the dimensionless dynamics of displacement $u$ and orientation $\theta$ reads
\begin{equation}\label{eq:dyn_1D}
	\dot u = \Pi \cos \theta - 2 u + u_{\rm op} ,
	\qquad
	\dot \theta = -\, \dot u \sin \theta + \sqrt{2D} \, \eta,
\end{equation}
where $\eta$ is a Gaussian white noise with unit variance and zero mean. From the definition of the operator force
\begin{equation}\label{eq:force_1D}
	F_{\rm op} = \langle u\rangle - u_{\rm op} ,
\end{equation}
we deduce that the extracted work $W$ [Eq.~(1) in main text] is given by
\begin{equation}\label{eq:work_1D}
	W = \int_0 ^{1/f_0} F_{\rm op} \dot{u}_{\rm op} dt = \int_0 ^{1/f_0} (u_{\rm op} - \langle u\rangle ) \dot{u}_{\rm op} dt .
\end{equation}
This expression is consistent with the standard definition of stochastic thermodynamics $W = \int_0^\tau \langle\partial U/\partial u_{\rm op}\rangle \, \dot u_{\rm op} dt$~\cite{Seifert2012}, where the potential $U = (1/2) (u-u_{\rm op})^2$ is associated with the force applied by the operator through the mobile structure; see Sec.~\ref{sec:exp} and Fig.~3(a) in the main text. In what follows, we discuss how to evaluate work $W_{\rm op}$ and power $P = W f_0$ in various regimes.

In the noiseless regime ($D=0$), there is a deterministic relation between $\theta$ and $u_{\rm op}$. The orientation dynamics $\dot\theta = - \dot u \sin\theta$ can be explicitly integrated in terms of $u_0=u(t=0)$ and $\theta_0=\theta(t=0)$ as
\begin{equation}
	u(t) = u_0 + \ln \bigg| \tan \frac{\theta_0}{2} \bigg| - \ln \bigg| \tan \frac{\theta(t)}{2} \bigg| ,
\end{equation}
which leads to the closed dynamics
\begin{equation}
	\dot\theta = - (\Pi \cos \theta - 2 u + u_{\rm op}) \sin\theta = - \frac{d V_{\rm eff}}{d\theta} ,
\end{equation}
where the effective potential $V_{\rm eff}$ reads
\begin{equation}\label{eq:veff}
\begin{aligned}
	V_{\rm eff}(\theta,t) &= - (\Pi/4) \cos(2\theta) - \cos\theta \ln | \tan (\theta/2) | + \ln |\sin\theta| - h_{\rm eff}(t) \cos \theta ,
	\\
	h_{\rm eff}(t) &= u_{\rm op}(t) - u_0 - \ln | \tan (\theta_0/2) | .
\end{aligned}
\end{equation}
Therefore, the effect of the operator displacement $u_{\rm op}$ can be regarded as an effective polarizing field $h_{\rm eff}$ applied to the orientation $\theta$. This effective potential picture breaks down for any amount of noise ($D>0)$.

\begin{figure*}
	\centering
	\includegraphics[width=\linewidth]{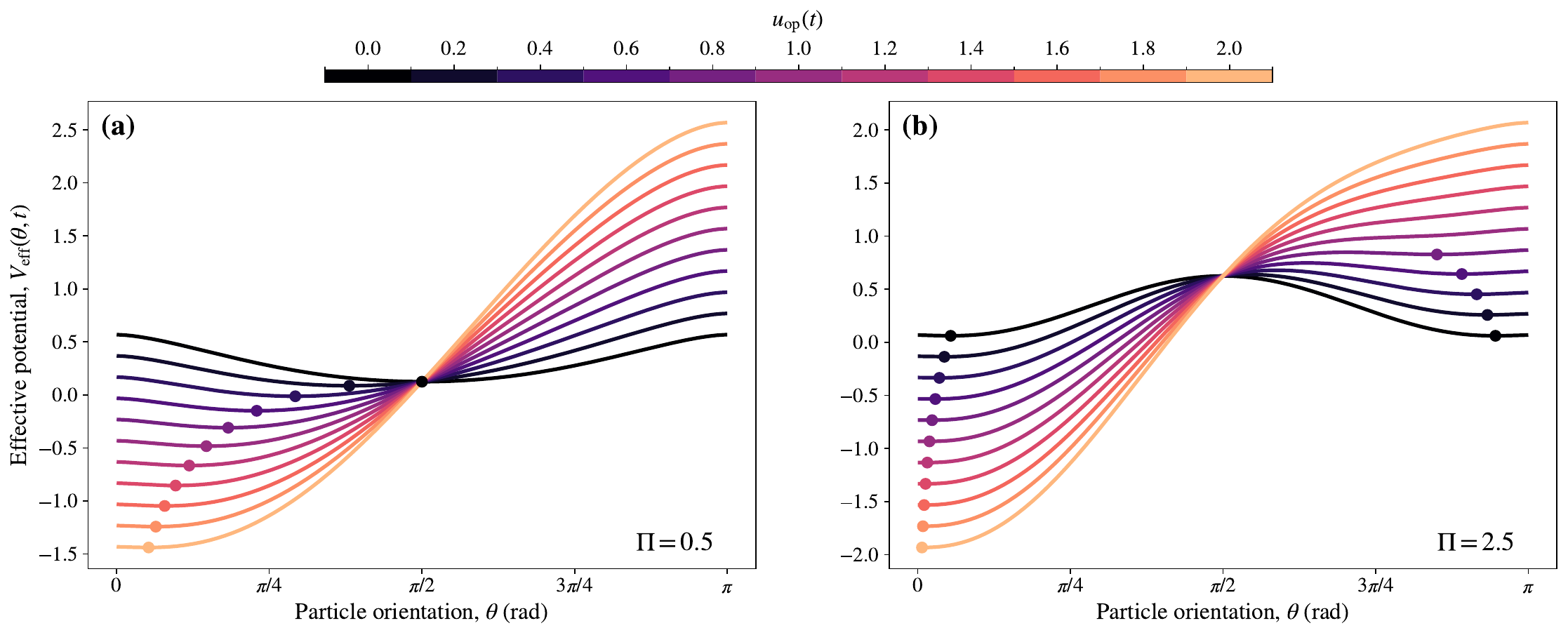}
	\caption{Effective potential $V_{\rm eff}(\theta, t)$ [Eq.~\eqref{eq:veff}] in the deterministic dynamics of orientation $\theta$ for a cycling drive $u_{\rm op}(t)$ with $(u_0=0,\theta_0=\pi/2)$.
	(a)~At low activity $\Pi$, there is always a single minimum in the potential; markers refer to locations of the global minimum $\theta^*(t) = \text{argmin}_\theta V_{\rm eff}(\theta,t)$.
	(b)~At high activity $\Pi$, the potential can feature several local minima.	
    }
	\label{fig:pot}
\end{figure*}

When $\Pi$ exceeds a threshold, $V_{\rm eff}$ features several minima, so $\theta^*(t) = {\rm argmin}_\theta V_{\rm eff}(\theta,t)$ can undergo transitions between hysteretic branches as $u_{\rm op}$ switches between compression and expansion [Fig.~\ref{fig:pot}]. Instead, there is a univocal relation between $\theta^*(t)$ and $u_{\rm op}$ at low activity: as discussed in the main text, such a relation precludes work extraction in this regime.

\subsection{Linear response theory}

We examine the regime where the operator displacement varies slowly compared with the relaxation time of the piston. At time $t$, we take as a reference the stationary state obtained by fixing the operator displacement to its instantaneous value $u_{\rm op}(t)$. The linear response of the average displacement to its previous values reads
\begin{equation}\label{eq:resp}
	\langle u(t)\rangle = \langle u\rangle_s + \int_{-\infty}^{t}dt'\, \Delta u_{\rm op}(t,t') \left.\frac{\delta\langle u(t)\rangle}{\delta u_{\rm op}(t')}\right|_{\Delta u_{\rm op}\to0} + O(\Delta u_{\rm op}^2),
	\qquad
	\Delta u_{\rm op}(t,t') = u_{\rm op}(t')-u_{\rm op}(t) ,
\end{equation}
where $\langle\cdot\rangle_s$ refers to a steady-state average. In all that follows, we omit the mention of small parameters (e.g., $\Delta u_{\rm op}\to0$) when writing down any responses and susceptibilities, so that they all implicitly correspond to linear perturbation. For a slowly varying protocol, we expand
\begin{equation}\label{eq:prot_slow}
	\Delta u_{\rm op}(t,t') = -(t-t')\dot u_{\rm op}(t) + {\cal O}(\ddot u_{\rm op}).
\end{equation}
To determine the stationary contribution $\langle u\rangle_s$, we first average the displacement dynamics [Eq.~\eqref{eq:dyn_1D}] at fixed $u_{\rm op}$:
\begin{equation} \label{eq:stationary_force_balance}
	0 = \Pi\langle\cos\theta\rangle_s - 2\langle u\rangle_s + u_{\rm op} .
\end{equation}
Note that the equations of motion [Eq.~\eqref{eq:dyn_1D}] have a simple translation symmetry: since $u$ and $u_{\rm op}$ enter the dynamics through the combination $-2u+u_{\rm op}$, their stationary distribution satisfies
\begin{equation} \label{eq:ps_symmetry}
	p_s(u+u_0,\theta;u_{\rm op}) = p_s(u,\theta;u_{\rm op}-2u_0) ,
\end{equation}
for an arbitrary constant $u_0$. After integration over $u$, this relation shows that the stationary angular distribution $p_s(\theta)=\int p_s(u,\theta)du$ (and hence $\langle\cos\theta\rangle_s$) is independent of $u_{\rm op}$. The inversion symmetry $\theta\to-\theta$ of the dynamics then gives $\langle\cos\theta\rangle_s=0$, so that
\begin{equation}
	\langle u\rangle_s=\frac{u_{\rm op}}{2}.
\end{equation}
Substituting the slow-protocol expansion [Eq.~\eqref{eq:prot_slow}] into Eq.~\eqref{eq:resp} yields
\begin{equation}\label{eq:zeta_resp}
	\langle u\rangle = \frac{u_{\rm op}}{2} - \zeta\dot u_{\rm op} + {\cal O}(\ddot u_{\rm op},\dot u_{\rm op}^{\,2}) ,
	\qquad
	\zeta = \int_0^\infty dt\,t \, \frac{\delta\langle u(t)\rangle}{\delta u_{\rm op}(0)} .
\end{equation}
The operator force $F_{\rm op} = u_{\rm op}-\langle u\rangle$ [Eq.~\eqref{eq:force_1D}] follows as
\begin{equation}
	F_{\rm op} = \frac{u_{\rm op}}{2} + \zeta\dot u_{\rm op} + {\cal O}(\ddot u_{\rm op},\dot u_{\rm op}^{\,2}),
\end{equation}
which is Eq.~(5) of the main text, from which we deduce the expression of extracted work $W$ [Eq.~\eqref{eq:work_1D}] to leading order:
\begin{equation}
	W = \int_0 ^{1/f_0} \left( \frac{u_{\rm op}}{2} + \zeta \dot{u} _{\rm op} \right) \dot{u}_{\rm op} dt = 4\zeta A^2 f_0 ,
\end{equation}
where we used $|\dot{u}_{\rm op} | = 2Af_0$, and the extracted power follows as $P = W f_0 = 4\zeta (Af_0)^2$.

We next show that $\zeta$, although defined through a time-dependent response [Eq.~\eqref{eq:zeta_resp}], can be related to a specific static susceptibility. In the presence of the polarizing field $h$, the dynamics reads
\begin{equation} \label{eq:eom_with_field}
	\dot u = \Pi\cos\theta-2u+u_{\rm op},
	\qquad
	\dot\theta = -(\dot u+h) \sin\theta+\sqrt{2D}\eta.
\end{equation}
The corresponding Fokker-Planck equation for the probability distribution $p$ is given in terms of the generator $\cal L$:
\begin{equation}
	\partial_t p = \mathcal Lp ,
	\qquad
	{\cal L} = - \partial_u \big[ \Pi\cos\theta-2u+u_{\rm op} \big] + \partial_\theta\big[ (\Pi\cos\theta-2u+u_{\rm op}+h) \sin\theta + D \partial_\theta\big] .
\end{equation}
The changes in the generator $({\cal L}_h, {\cal L}_{u_{\rm op}})$ respectively associated with variations of $h$ and $u_{\rm op}$ are
\begin{equation} \label{eq:op_identity_1}
	\mathcal L_h = \partial_\theta(\sin\theta) ,
	\qquad
	\mathcal L_{u_{\rm op}} = -\partial_u + \partial_\theta (\sin\theta) ,
\end{equation}
so their difference reads
\begin{equation}\label{eq:op_diff}
	\mathcal L_h-\mathcal L_{u_{\rm op}} = \partial_u .
\end{equation}
For an arbitrary observable $g(u,\theta)$, linear response around the stationary distribution $p_s$ gives~\cite{Baiesi2013}
\begin{equation}\label{eq:der_1}
	\frac{\delta\langle g(t)\rangle}{\delta a(0)} = \frac{\delta}{\delta a(0)} \int du\,d\theta\,
g\, e^{\mathcal Lt} p_s = \int du\,d\theta\,
g\, e^{\mathcal Lt}\mathcal L_a p_s,
\end{equation}
where $a$ denotes either $h$ or $u_{\rm op}$, and using Eq~\eqref{eq:op_diff}, the difference between the two responses becomes
\begin{equation}
	\frac{\delta\langle g(t)\rangle}{\delta h(0)} - \frac{\delta\langle g(t)\rangle}{\delta u_{\rm op}(0)} = \int du\,d\theta\, g\,e^{\mathcal Lt}\partial_u p_s.
\end{equation}
Given the symmetry of the stationary distribution $p_s$ [Eq.~\eqref{eq:ps_symmetry}], it follows that differentiating $p_s$ with respect to $u_0$ at $u_0=0$ yields $\partial_u p_s = -2\partial_{u_{\rm op}}p_s$, so that
\begin{equation}
	\frac{\delta\langle g(t)\rangle}{\delta h(0)} - \frac{\delta\langle g(t)\rangle}{\delta u_{\rm op}(0)} = -2 \int du\,d\theta\, g\,e^{\mathcal Lt} \partial_{u_{\rm op}}p_s .
\end{equation}
The last term can itself be expressed in terms of the response to $u_{\rm op}$. The stationary distribution obeys $\mathcal Lp_s=0$. Differentiating this equation with respect to $u_{\rm op}$ yields $\mathcal L\partial_{u_{\rm op}}p_s = -\mathcal L_{u_{\rm op}}p_s$.
It follows that
\begin{equation}
	\frac{d}{dt} \left[ e^{\mathcal Lt} \partial_{u_{\rm op}}p_s \right] = - e^{\mathcal Lt} \mathcal L_{u_{\rm op}}p_s.
\end{equation}
Since $\partial_{u_{\rm op}}p_s$ has zero normalization, its propagation relaxes to zero at long times. Integrating the previous equation from $t$ to infinity therefore gives
\begin{equation}
	e^{\mathcal Lt} \partial_{u_{\rm op}}p_s = \int_t^\infty dt'\, e^{\mathcal Lt'} \mathcal L_{u_{\rm op}}p_s .
\end{equation}
Multiplying by $g$ and integrating over phase space, we obtain
\begin{equation}
\int du\,d\theta\, g\,e^{\mathcal Lt} \partial_{u_{\rm op}}p_s = \int_t^\infty dt'\, \frac{\delta\langle g(t')\rangle} {\delta u_{\rm op}(0)} ,
\end{equation}
which is known as the Agarwal-Kubo relation~\cite{Baiesi2013}. Combining the above results gives the response identity
\begin{equation}\label{eq:resp_diff}
	\frac{\delta\langle g(t)\rangle}{\delta h(0)} - \frac{\delta\langle g(t)\rangle}{\delta u_{\rm op}(0)} = -2 \int_t^\infty dt'\,
\frac{\delta\langle g(t')\rangle} {\delta u_{\rm op}(0)}.
\end{equation}
We now set $g=u$ and integrate Eq.~\eqref{eq:resp_diff} over $t$ from zero to infinity. The integrated response on the left-hand side gives the corresponding static susceptibilities, while the right-hand side can be evaluated by exchanging the order of integration:
\begin{equation}
\begin{aligned}
	\frac{\partial\langle u\rangle_s}{\partial h} - \frac{\partial\langle u\rangle_s}{\partial u_{\rm op}} &= -2 \int_0^\infty dt \int_t^\infty dt'\,\frac{\delta\langle u(t')\rangle}{\delta u_{\rm op}(0)}
	\\
	&= -2 \int_0^\infty dt'\,\int_0^{t'}dt\,\frac{\delta\langle u(t')\rangle}{\delta u_{\rm op}(0)}
	\\
	&= -2 \int_0^\infty dt'\,t'\frac{\delta\langle u(t')\rangle}{\delta u_{\rm op}(0)} = - 2 \zeta ,
\end{aligned}
\end{equation}
where we have used the definition of $\zeta$ [Eq.~\eqref{eq:zeta_resp}]. Combining the stationary force balance [Eq.~\eqref{eq:stationary_force_balance}] with the independence of $\langle\cos\theta\rangle_s$ from $u_{\rm op}$, we deduce
\begin{equation}
	\frac{\partial\langle u\rangle_s} {\partial u_{\rm op}} = \frac12 ,
	\qquad
	\frac{\partial\langle u\rangle_s}{\partial h} = \frac\Pi2 \frac{\partial\langle\cos\theta\rangle_s}{\partial h} ,
\end{equation}
yielding
\begin{equation}\label{eq:zeta}
	\zeta = \frac14 \left( 1-\Pi \frac{\partial\langle\cos\theta\rangle_s}{\partial h} \right) ,
\end{equation}
which is Eq.~(6) of the main text.

In a similar way, we compute $\zeta$ in the case without self-alignment, for which the dynamics reads
\begin{equation}
    \dot{u} = \Pi \cos \theta - 2u + u_{\rm{op} } ,
    \qquad \dot{\theta} = -h \sin \theta + \sqrt{2D} \eta 
    .
\end{equation}
Then, the operator identity [Eq.~\eqref{eq:op_identity_1}] changes to 
\begin{equation}
    -\mathcal{L}_{u_{\rm{op}}} = \partial_u .
\end{equation}
Using the same logic as in Eqs.~(\ref{eq:der_1}-\ref{eq:resp_diff}), we arrive at
\begin{equation}
    \frac{\delta \langle g(t)\rangle}{\delta u_{\rm{op}}(0)} = 2 \int_{t}^\infty dt' \frac{\delta \langle g(t') \rangle}{\delta u_{\rm{op}}(0)} .
\end{equation}
Setting $g=u$ and integrating the above over $t$ from zero to infinity, we deduce
\begin{equation}
    \zeta = \frac{1}{2} \frac{\partial \langle u \rangle_s}{\partial u_{\rm{op}}} = \frac{1}{4} ,
\end{equation}
showing that the response coefficient $\zeta$ becomes independent of activity $\Pi$ and noise $D$ in the absence of self-alignment.


\subsection{Extracted power}

Our aim is to derive some analytical prediction of the power $P=f_0 W$. To this end, we note that the force exerted by the operator can be written as
\begin{equation}
	F_{\rm op} = u_{\rm op} -\langle u\rangle = \frac{1}{2} ( \langle\dot u\rangle - \Pi \langle\cos\theta\rangle + u_{\rm op} ) ,
\end{equation}
and substituting this result into the expression of the work [Eq.~\eqref{eq:work_1D}] yields the decomposition
\begin{equation}
	W = W_{\rm el} + W_{\rm ac} + W_{\rm fr} ,
	\quad
    W_{\rm el} = \frac{1}{2} \int u_{\rm op} \, d u_{\rm op} ,
	\quad
    W_{\rm fr} = \frac{1}{2} \int \langle \dot u \rangle \, d u_{\rm op} ,
	\quad
	W_{\rm ac} = - \frac{\Pi}{2} \int \langle \cos\theta \rangle \, d u_{\rm op} ,
\end{equation}
where $W_{\rm el} = 0$ due to integration on a closed path. To evaluate $(W_{\rm ac}, W_{\rm fr})$, we proceed by focusing on the regime of low activity $\Pi$. In the passive limit ($\Pi=0$), the displacement $u$ decouples from the orientation $\theta$, and relaxes to $u = u_{\rm op}/2$. Therefore, we approximate $\dot u  = \dot u_{\rm op}/2$. The operator displacement velocity is given by $\dot u_{\rm op} = 2 A f_0$ during compression $t\in[0,1/(2f_0)]$, and $\dot u_{\rm op} = - 2 A f_0$ during expansion $t\in[1/(2f_0),1/f_0]$. Moreover, the trajectory of $\langle \cos \theta \rangle$ takes values that are equal and opposite during compression and expansion, yielding
\begin{equation}\label{eq:wfr}
	W_{\rm fr} = \frac 14 \int_0^{1/f_0} (\dot u_{\rm op})^2 d t = A^2 f_0 ,
	\quad
	W_{\rm ac} = - \frac{\Pi}{2} \int \langle \cos\theta \rangle \, \dot u_{\rm op} dt = - 2 \Pi A f_0 \int_0^{1/(2f_0)} \langle \cos \theta \rangle dt .
\end{equation}
In what follows, we propose two distinct procedures to evaluate $\langle\cos\theta\rangle$ at small and large $D$. In each regime, we deduce that the extracted power takes the form
\begin{equation}\label{eq:power}
	P = (A f_0)^2 \bar P(A, \Pi/f_0, D/f_0) ,
\end{equation}
so the phase boundary $P=0$ follows a master curve in terms of $(D/f_0,\Pi/f_0)$, as detailed in the main text.


\subsubsection{Large noise approximation}

The typical relaxation rate is set by $D$. Then, at large $D$, we assume that the protocol is so slow compared to spontaneous relaxation that it effectively drives the system through a series of steady state. In this regime, we approximate
\begin{equation}\label{eq:wac}
	\langle\cos\theta\rangle \approx \langle\cos\theta\rangle_s = \int p_s(\theta) \cos\theta d\theta ,
\end{equation}
where the stationary distribution $p_s$ follows from the Fokker-Planck equation associated with the Langevin dynamics for $\theta$. The low-$\Pi$ approximation ($\dot u = \dot u_{\rm op}/2$) used in the evaluation of the work [Eq.~\eqref{eq:wfr}] leads us to simplify the dynamics of $\theta$ [Eq.~\eqref{eq:dyn_1D}] as
\begin{equation}\label{eq:approx}
	\dot\theta = - A f_0 \, \sin \theta + \sqrt{2D} \, \eta ,
\end{equation}
where we have used $\dot u_{\rm op}/2 = A f_0$ during compression, from which we deduce the Fokker-Planck equation
\begin{equation}
	\partial_t p = \partial_\theta \Big[ A f_0 p \sin\theta + D \partial_\theta p \Big] ,
\end{equation}
yielding
\begin{equation}\label{eq:dist}
	p_s = \frac{1}{Z} e^{A f_0\cos\theta/D } ,
	\qquad
	Z = \int e^{A f_0\cos\theta/D} d\theta .
\end{equation}
Substituting the distribution $p_s$ [Eq.~\eqref{eq:dist}] into $\langle\cos\theta\rangle$ [Eq.~\eqref{eq:wac}], we deduce the expression of $W_{\rm ac}$ [Eq.~\eqref{eq:wfr}] as
\begin{equation}\label{eq:wac_gg}
	W_{\rm ac} = - \Pi A \langle \cos\theta \rangle = - \Pi A \, \frac{I_1(Af_0/D)}{I_0(Af_0/D)} ,
	\qquad
	I_n(z) = \int \cos(n\theta) e^{z \cos\theta} d\theta ,
\end{equation}
where $I_n$ is the modified Bessel functions of the first kind. It follows that in the large noise limit, the power takes the form of Eq.~\eqref{eq:power}, where $\bar P = \bar P_\gg$ reads
\begin{equation}\label{eq:pgg}
	\bar P_\gg = 1 - \frac{\Pi}{A f_0} \frac{I_1(Af_0/D)}{I_0(Af_0/D)} .
\end{equation}
We now show that this result is consistent with the linear response prediction $P = 4(Af_0)^2 \zeta$, where the linear response coefficient obeys $4\zeta = 1 - \Pi \frac{\partial \langle\cos\theta\rangle_s}{\partial h} $ [Eq.~\eqref{eq:zeta}]. Here, $\langle\cos\theta\rangle_s$ corresponds to the stationary average in the unperturbed dynamics, namely for $\dot u_{\rm op}=0$, which features a complex dependence on $(\Pi,D,h)$~\cite{Baconnier2025_prl, Baconnier2025_pre}. To leading order in $\Pi$, the dynamics for $\theta$ and its corresponding stationary distribution reads
\begin{equation}
	\dot\theta ={\color{red}-} h \sin\theta + \sqrt{2D} \eta ,
	\qquad
	p_s = \frac 1 Z e^{h\cos\theta/D} ,
	\qquad
	Z = \int e^{h \cos\theta/D} d\theta ,
\end{equation}
yielding
\begin{equation}\label{eq:zeta}
	\zeta = \frac{1}{4} \left( 1 - \frac{\Pi}{2D} \right).
\end{equation}
Expanding $P_\gg$ [Eq.~\eqref{eq:pgg}] at large $D/f_0$, we obtain
\begin{equation}
	\bar P_\gg = 1 - \Pi/(2D) + {\cal O}(f_0/D) ,
\end{equation}
in agreement with the linear response prediction [Eq.~\eqref{eq:zeta}]. The corresponding phase boundary of vanishing power ($\Pi=2D$) is monotonic, so linear response is inadequate to capture any re-entrance in the phase diagram.


\begin{figure*}
	\centering
	\includegraphics[width=\linewidth]{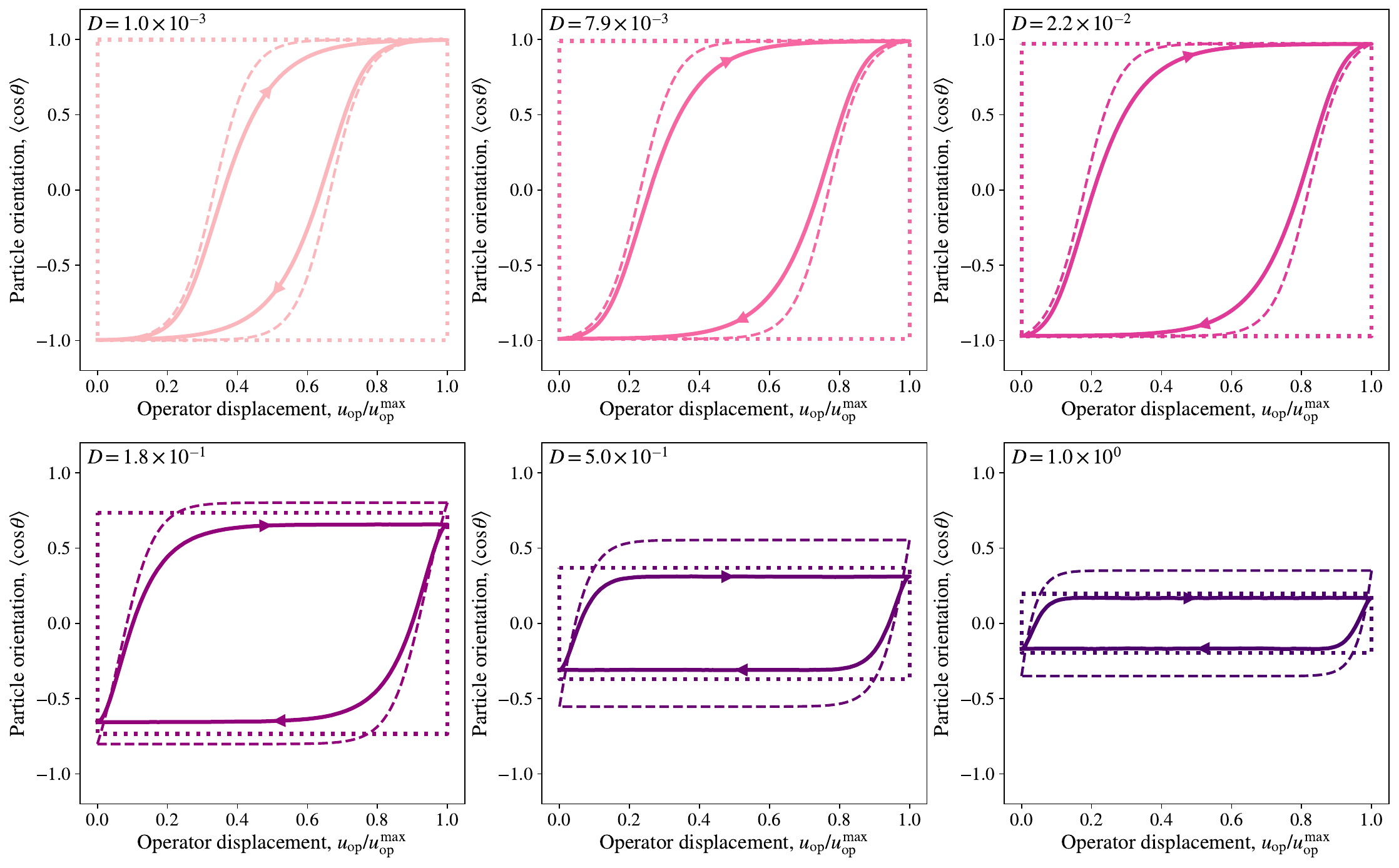}
	\caption{Mechanical cycles of the angular moment $q=\langle\cos\theta\rangle$ as a function of the operator displacement $u_{\rm op}$ for different values of noise $D$: dashed lines refer to analytical predictions at small noise [Eq.~\eqref{eq:q}], and dotted lines to the linear response regime at large noise, for which plateau values are given by $\langle\cos\theta\rangle \approx \langle\cos\theta\rangle_s$ [Eq.~\eqref{eq:wac_gg}].
    Parameters: $\Pi=1$, $f_0=0.02$, $A=20$.
	}
	\label{fig0}
\end{figure*}

\subsubsection{Small noise approximation}

The stochastic It\^o dynamics of the angular moment $n=\cos\theta$ follows from  the dynamics of orientation $\theta$ [Eq.~\eqref{eq:approx}] using It\^o's lemma~\cite{Gardiner} as
\begin{equation}\label{eq:mom}
	\dot n = - D \, n + A f_0 (1-n^2) + \sqrt{2D\,(1-n^2)} \, \eta .
\end{equation}
At small $D$, we use the heuristic closure $\langle n^2\rangle = \langle n \rangle^2$ to obtain a closed dynamics for $q = \langle\cos\theta\rangle$ by averaging the moment dynamics [Eq.~\eqref{eq:mom}]:
\begin{equation}
	\dot q = A f_0(1-q^2) - D \, q ,
\end{equation}
whose solution can be written for a given $q_0 = q(t=0)$ as
\begin{equation}\label{eq:q}
\begin{aligned}
	&q(t) = C\bigg(\frac{A f_0}{D}\bigg) \, \tanh\bigg[ {\rm argtanh} \bigg( \frac{ D/(2Af_0) + q_0}{C(A f_0/D)} \bigg) + Af_0 t \, C \bigg(\frac{A f_0}{D}\bigg) \bigg] - \frac{D}{2 Af_0} ,
	\\
	&C\bigg(\frac{A f_0}{D}\bigg) = \sqrt{1 + \bigg(\frac{D}{2 A f_0}\bigg)^2} .
\end{aligned}
\end{equation}
Substituting $q(t)$ [Eq.~\eqref{eq:q}] into $W_{\rm ac}$ [Eq.~\eqref{eq:wfr}], it follows that the power $P$ takes the form in Eq.~\eqref{eq:power}, where $\bar P = \bar P_\ll$ reads
\begin{equation}\label{eq:pll}
	\bar P_\ll = 1 + \frac{\Pi}{A^2 f_0} \bigg( \frac{D}{2f_0} + \ln \frac{1 - (D/Af_0) q_0 -q_0 ^2}{1 + (D/Af_0) q_0 - q_0 ^2}\bigg). 
\end{equation}
At the beginning and end of compression, $q(t)$ obeys the symmetry $q(t=0) = - q(t=1/(2 f_0))$, yielding
\begin{equation}
	q_0 = \frac{C(A f_0/D)}{\tanh( (A/2)  C(A f_0/D) )} - \sqrt{ \bigg[ \frac{C(A f_0/D)}{\tanh( (A/2)  C(A f_0/D) )} \bigg]^2 - 1 } .
\end{equation}
Therefore, $q_0$ and $\bar P_\ll$ are functions of $(A,\Pi/f_0, D/f_0)$. In practice, comparing our prediction for $q(t)$ [Eq.~\eqref{eq:q}] during expansion and compression ($u_{\rm op}=\pm 2 A f_0 t$) with results from numerical simulations shows that our approach indeed captures well the emergent mechanical cycles $(q, u_{\rm op})$ at small $D$ [Fig.~\ref{fig0}].


\subsubsection{Phase diagrams and master curve}

We now examine the dependence of the power $P$ in terms of noise $D$ and activity $\Pi$. Based on our prediction at small $D$ [$\bar P_\ll$, Eq.~\eqref{eq:pgg}] and large $D$ [$\bar P_\gg$, Eq.~\eqref{eq:pll}], the boundary in the phase diagram $(D,\Pi)$ that distinguishes the regimes of extracted ($P<0$) and supplied ($P>0$) power follows from the condition of vanishing power ($P=0$). Given that $(\bar P_\ll, \bar P_\gg)$ are both functions of $(A,\Pi/f_0,D/f_0)$, we deduce that phase boundaries follow a master curve for all $f_0$ at a given $A$. In practice, comparison with boundaries obtained from numerical simulations shows that the agreement with our predictions $(\bar P_\ll, \bar P_\gg)$ gets better as $A$ increases [Fig.~\ref{fig2}].

\begin{figure*}
	\centering
	\includegraphics[width=\linewidth]{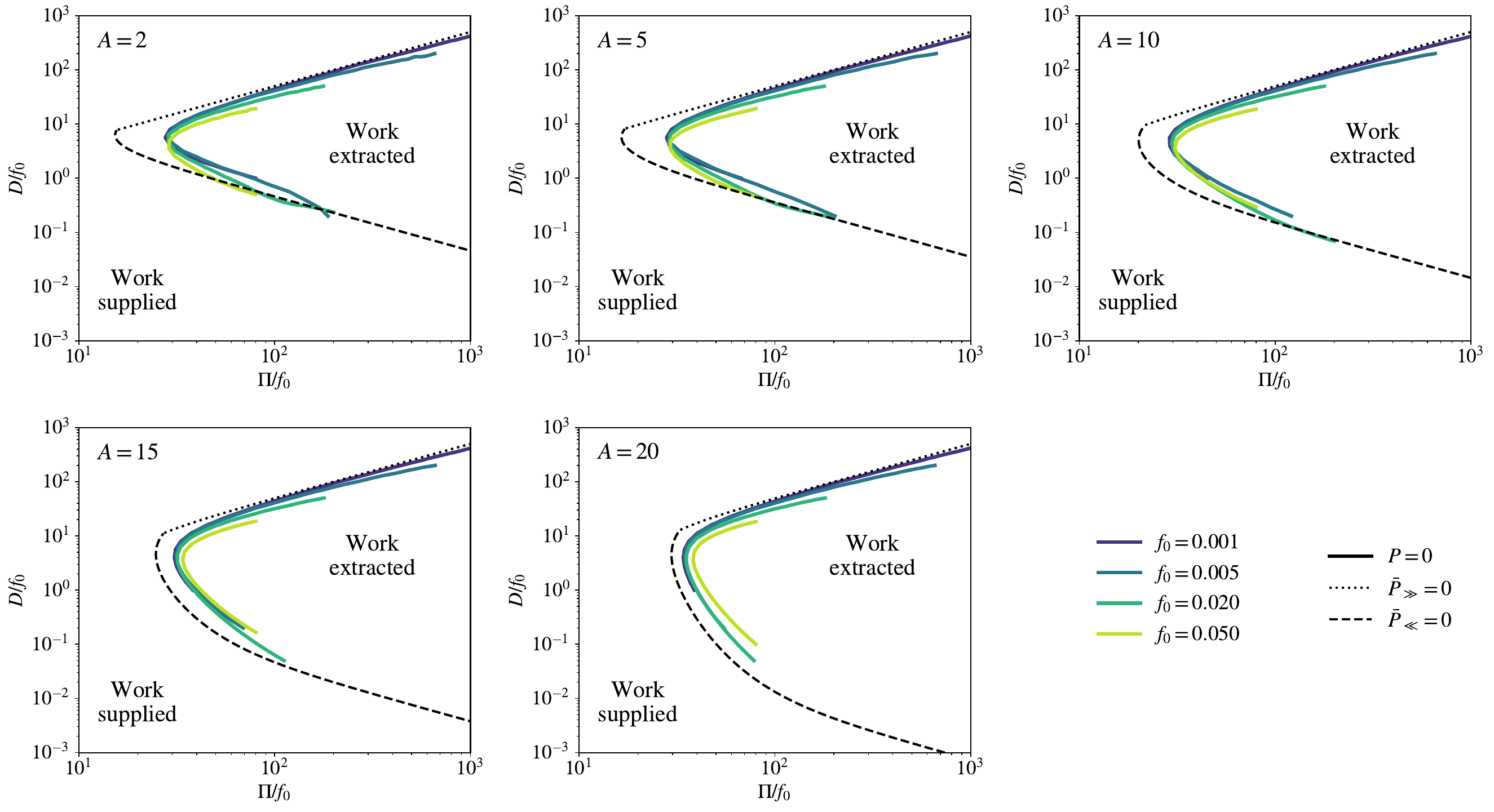}
	\caption{Phase boundaries of vanishing power ($P=0$) delineate for each value of operator displacement range $A$ a regime of work extraction ($P<0$) at large $\Pi/f_0$ and intermediate $D/f_0$: colored solid lines for simulation results, black dashed and dotted lines for analytical predictions; see $\bar P_\gg$ in Eq.~\eqref{eq:pgg}, and $\bar P_\ll$ in Eq.~\eqref{eq:pll}.
	}
	\label{fig2}
\end{figure*}


\subsection{Numerical simulations and evaluation of linear response}

Here, we provide details about the numerical simulations of the one-dimensional active piston with a single hexbug used in Figs.~3 and 4 of the main text. We integrate Eq.~\eqref{eq:eom_with_field} using the Euler-Maruyama scheme with time step $\Delta t=0.01$. Denoting $t_n=n\Delta t$, the discretized dynamics reads
\begin{equation}
\label{eq:euler_maruyama}
\begin{aligned}
u(t_{n+1})
&=
u(t_n)
+\Delta t
\left[
\Pi\cos\theta(t_n)
-2u(t_n)
+u_{\rm op}(t_n)
\right],
\\
\theta(t_{n+1})
&=
\theta(t_n)
-\Delta t
\left[
\dot u(t_n)+h
\right]
\sin\theta(t_n)
+\xi(t_n),
\end{aligned}
\end{equation}
where $\xi(t_n)$ is an independent Gaussian random variable satisfying
\begin{equation}
\langle \xi(t_n)\rangle=0,
\qquad
\langle \xi(t_n)\xi(t_{n'})\rangle
=
2D\Delta t\,\delta_{nn'}.
\end{equation}
To compute the average power, we simulate ten complete driving cycles. For each independent realization, we set $u(0)=0$ and draw $\theta(0)$ uniformly from $[0,2\pi)$. The first five cycles are discarded as transients, and the power is obtained by time averaging over the remaining five cycles. The resulting quantity is then averaged over $N_{\rm ens}=10^4$ independent realizations.

In Figs.~4(a) and 4(c) of the main text, the coefficient $\zeta$ is evaluated using Eq.~(6) of the main text [Eq.~\eqref{eq:zeta}], which expresses $\zeta$ in terms of the static susceptibility $\partial_h\langle\cos\theta\rangle_s$. We determine this susceptibility numerically as follows. For each realization $k\in\{1,\ldots,N_{\rm ens}\}$, we first generate and store a noise sequence
\begin{equation}
\left\{
\xi^k(t_n)
\right\}_{n=0}^{M_{\rm th}+M_{\rm meas}},
\end{equation}
where $M_{\rm th}$ and $M_{\rm meas}$ denote the numbers of thermalization and measurement steps, respectively. We use
\begin{equation}
M_{\rm th}=10^3,
\qquad
M_{\rm meas}=6\times10^5.
\end{equation}
The same noise sequence is used for the trajectories at finite $h$ and for the corresponding reference trajectory at $h=0$. This common-noise protocol substantially reduces the statistical fluctuations in their difference. For each value of the field
\begin{equation}
h\in
\left\{
10^{-4},\,10^{-3},\,2\times10^{-3},\,
5\times10^{-3},\,10^{-2}
\right\},
\end{equation}
we evolve Eq.~\eqref{eq:euler_maruyama} using the time-dependent field
\begin{equation}
h_n=
\begin{cases}
0,
& 0\le n<M_{\rm th},
\\[2mm]
h,
& M_{\rm th}\le n<M_{\rm th}+M_{\rm meas}.
\end{cases}
\label{eq:h_protocol}
\end{equation}
Thus, the system is first thermalized at $h=0$, and the field $h$ is then switched on only during the measurement stage.

To suppress the transient following the switching of $h$, we discard the first half of the measurement interval. We then evaluate the field-induced change in the stationary orientational polarization as
\begin{equation}
\label{eq:delta_cos_response}
\delta\langle\cos\theta\rangle(h)
=
\frac{2}{M_{\rm meas}N_{\rm ens}}
\sum_{k=1}^{N_{\rm ens}}
\sum_{n=M_{\rm th}+M_{\rm meas}/2}^{M_{\rm th}+M_{\rm meas}-1}
\left[
\cos\theta^k(t_n;h)
-
\cos\theta^k(t_n;0)
\right].
\end{equation}
Here, $\theta^k(t_n;h)$ denotes the orientation obtained for the $k$th noise realization with the field protocol in Eq.~\eqref{eq:h_protocol}, while $\theta^k(t_n;0)$ denotes the corresponding reference trajectory with $h=0$ throughout. Importantly, the two trajectories are generated using the same initial condition and the same noise realization.

\begin{figure*}
	\centering
	\includegraphics[width=\linewidth]{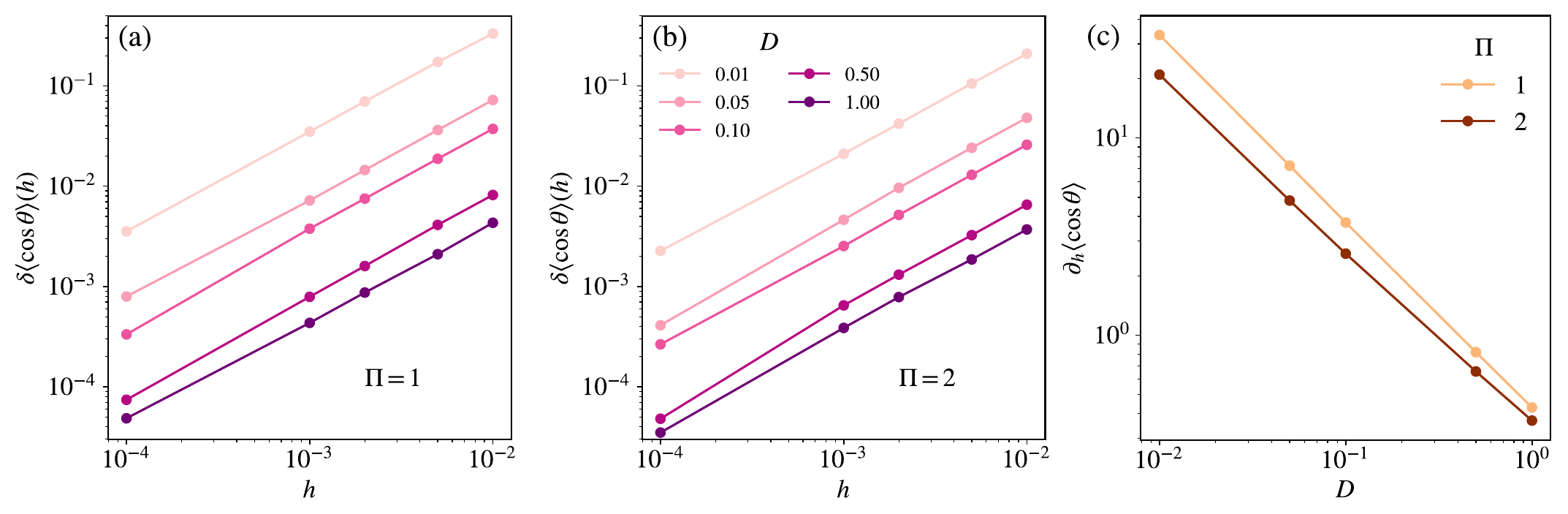}
	\caption{
    Evaluation of linear response $\delta \langle \cos \theta \rangle (h) $ [Eq.~\eqref{eq:delta_cos_response}] with different noise $D$ for (a)~$\Pi = 1$, and (b)~$\Pi = 2$.
    (c)~Estimation of $\partial_h \langle \cos \theta \rangle$  by a linear regression of $\delta \langle \cos \theta \rangle (h)$ against $h$ for $\Pi \in\{ 1,2\}$. 
	}
	\label{fig7}
\end{figure*}

Finally, within the linear-response regime, we can use the relation
\begin{equation}
\delta\langle\cos\theta\rangle(h)
=
\left.
\frac{\partial\langle\cos\theta\rangle_s}{\partial h}
\right|_{h=0}
h
+O(h^2).
\end{equation}
We therefore estimate $\partial_h\langle\cos\theta\rangle_s|_{h=0}$ from the slope of a linear regression of $\delta\langle\cos\theta\rangle(h)$ against $h$. In Fig.~\ref{fig7}, we plot $\delta \langle \cos \theta \rangle (h)$ with different values of $D$ and $\Pi$. For each $D$ and $\Pi$, $\delta \langle \cos \theta \rangle (h)$ is a linear function of $h$, and the linear regression shown in panel (c) corresponds to $\partial_h \langle \cos \theta \rangle$.

\end{document}